\documentstyle[12pt,epsfig]{article}

\newcommand{\be}{\begin{equation}}
\newcommand{\ee}{\end{equation}}
\newcommand{\bea}{\begin{eqnarray}}
\newcommand{\eea}{\end{eqnarray}}

\title{ \hfill $\mbox{\small{
$\stackrel{\rm\textstyle SU{-}4240{-}691\quad}
{\rm \textstyle \quad}
$}}$ \\[1truecm]
The Renormalization Group and its Finite Lattice Approximations}

\author{\\\\ Angelo Cacciuto \thanks{\tt cacciuto@epictet.npac.syr.edu},
  Eric Gregory\thanks{\tt gregory@npac.syr.edu}
 and Alex Travesset\thanks{\tt alex@suhep.phy.syr.edu}
\\\\ Physics Department, Syracuse University,\\
Syracuse, NY 13244-1130, USA \\ }

\date{}

\begin{document}

\begin{titlepage}
\maketitle
\begin{abstract}

We investigate finite lattice approximations to the Wilson Renormalization
Group in models of unconstrained spins. We discuss first the properties 
of the Renormalization Group Transformation (RGT) that control the 
accuracy of this type of approximations and explain different methods 
and techniques to practically identify them. We also discuss
how to determine the anomalous dimension of the field. We apply
our considerations to a linear sigma model in two dimensions 
in the domain of attraction of the Ising Fixed Point using a
Bell-Wilson RGT. We are able to identify optimal RGTs which allow 
accurate computations of quantities such as critical exponents, fixed 
point couplings and eigenvectors with modest statistics. We finally discuss 
the advantages and limitations of this type of approach.

\end{abstract}
\end{titlepage}

\section{Introduction}\label{sect__inflat}

\subsection{Scope and organization of the paper}

Finite lattice approximations provide an alternative and, we think, 
largely unexplored approach to the Renormalization Group (RG) \cite{WK}.
The advantage of a direct RG approach lies in its generality 
as not only does it provide, in principle, a direct and complete calculation 
of the critical exponents of the model, but also further valuable information 
such as the fixed point Hamiltonian and the renormalized trajectory.
In practical calculations, however, the apparent simplicity of the method 
is usually complicated because one needs to introduce further 
approximations (truncation in the operator basis, too few iterations
of the RG, etc.). In addition there are errors inherent to the approach
(finite size effects, statistical errors in numerical simulations, etc.).
Models of unconstrained spins present, in addition, the problem of
determining the rescaling of the field.

These difficulties may be overcome if a detailed knowledge of the 
properties of the renormalization group transformation used are
available. We discuss  these properties in some detail and 
explain different criteria, some already known in the literature,
to single out the best renormalization group transformation and
determine the anomalous dimension of the field.

This paper is organized as follows. In Sect.~\ref{sect__preparat} we
discuss first some general aspects of the Renormalization Group (RG) 
and the Monte Carlo Renormalization Group (MCRG) technique. 
We then describe the parameterizations of the effective Hamiltonians
generated along the RG flow more appropriate for the kind of approximation
to be applied. We finally introduce the particular RGT to be used in 
this paper and we discuss the physical meaning of the parameters involved.

In Sect.~\ref{sect__finlat} we address the problems and subtleties 
that appear in the finite lattice approximations of the RG. We first 
discuss the theoretical aspects and then explain different methods 
and techniques to be used in practice. We also discuss on the different 
sources of possible errors involved.

The technical details of the algorithms used, as well as the description
of the actual parameters of the simulation, are presented in 
Sect.~\ref{sect__numdet}.

In Sect.~\ref{sect__results} we present in detail the numerical
results concerning short-rangeness of the effective interactions
generated along the flow, optimization of the transformation, determination
of the anomalous dimension, calculation of the critical exponents and
the estimate of the different errors involved in the calculation.

The results presented in Sect.~\ref{sect__results} are rather
technical, so we summarize them in the first
part of the conclusions. In the second part we discuss further 
improvements and future outlook. Some technical issues are discussed 
in the appendices.

\section{The RG} \label{sect__preparat}

\subsection{Some generalities of the RG} 

The idea of the RG\cite{WK} is simple; average out short-distance modes
leaving large distance properties intact. Let us assume that our model
of interest is described by the Hamiltonian (so called canonical surface
in the RG language)
\be\label{def_model}
{\cal H}^0[\phi]=\sum_{\alpha} K_{\alpha}^{0} O_{\alpha}(\phi) \ ,
\ee
where $\{K_{\beta}^0\}_{\beta=1,\cdots}$ are the couplings and
$\{ O_\alpha \}_{\alpha=1,\cdots}$ an arbitrarily chosen basis of operators.
The integration of short-distance modes is implemented by  
a transformation ${\cal R}$, the Renormalization Group Transformation (RGT),
whose result is to define a new effective interaction
${\cal H}^1$ which we parameterize with the same basis of operators as 
the canonical surface. The net effect of the RGT is a transformation in 
the space of all possible couplings of the model,
\be\label{RGT_def}
K_{\alpha}^1={\cal R}_{\alpha}(\{K_{\beta}^0\}) \ .
\ee
Iteratively applying Eq.~\ref{RGT_def} we generate a sequence of points
in coupling space that define a trajectory. We can visualize, then, 
the effect of a particular RGT as trajectories flowing in coupling space.
Usually, these trajectories converge to a special point, 
say $\{K_{\alpha}^{*}\}_{\alpha=1,\cdots}$, a
fixed point(FP). It is a fixed point of the RGT Eq.~\ref{RGT_def},
\be\label{RGT_FP_def}
K_{\alpha}^*={\cal R}_{\alpha}(\{K_{\beta}^* \}) \ .
\ee
The importance of a FP speaks for itself. The universal properties of all
Hamiltonians lying in trajectories converging to this FP are completely
characterized by the the RG flow in an infinitesimal vicinity around it
\cite{WK}; all the critical exponents, for example,
just follow from diagonalizing the matrix
\be\label{T_def}
{\rm T}_{\alpha \beta}=\left.\frac{\partial{\cal R}_{\alpha}(K)}{\partial 
K_{\beta}} \right|_{\{K_{\gamma}^*\}} \ .
\ee

In this way, the RG provides a completely general and exact prescription 
to compute the critical properties of any model. 

\subsection{The MCRG method}

The exact ${\rm T}$ matrix is a difficult object to compute. There are
a number of approaches available, but as our interest lies in applying 
numerical techniques,  we seek an expression for the {\rm T}
matrix as a function of expectation values of operators. Remarkably, this can 
be achieved \cite{SWEN1}. At the $(i+1)$th iteration of the 
RGT Eq.~\ref{RGT_def}
one can easily prove \cite{SWEN1}
\be\label{der_coup}
\left.\frac{\partial{\cal R}_{\alpha}(K)}{\partial 
K_{\beta}} \right|_{\{K_{\gamma}^i\}}=\sum_{\gamma} {\rm H}^{-1}_{\alpha \gamma}
{\rm M}_{\gamma \beta} \ ,
\ee
with $\gamma$ running over all possible operators and
\bea\label{def_matrices}
{\rm M}_{\gamma \beta}&=&\langle O_{\gamma}^{(i+1)} O_{\beta}^{(i)} \rangle
-\langle O_{\gamma}^{(i+1)} \rangle \langle O_{\beta}^{(i)} \rangle \ ,
\nonumber\\
{\rm H}_{\gamma \beta}&=&\langle O_{\gamma}^{(i+1)} O_{\beta}^{(i+1)} \rangle
-\langle O_{\gamma}^{(i+1)} \rangle \langle O_{\beta}^{(i+1)} \rangle \ .
\eea
These are expectation values of operators at the $i$th and $(i+1)$th 
iteration of the RGT, but can be computed using ${\cal H}^k$ with $k \leq i$.
If ${\cal H}^i$ is the 
FP-Hamiltonian, which obviously is the case if $i=\infty$, Eq.~\ref{der_coup} 
is the exact ${\rm T}$ matrix, as defined in Eq.~\ref{T_def}.

With the very same effort, the couplings of the effective interaction
${\cal  H}^{(i+1)}$ can also be computed \cite{GONZ}. The formula is
\footnote{This formula applies for linear sigma models.
Other models require some adaptations.}
\be\label{def_coup}
K_{\beta}^{i+1}=\frac{1}{d_{\beta}} \sum_{\alpha} {\rm H}^{-1}_{\beta \alpha} 
d_{\alpha} A_{\alpha} \ ,
\ee
where the ${\rm H}$ matrix is defined in Eq.~\ref{def_matrices}, 
$A_{\alpha}=\langle O_{\alpha}^{i+1} \rangle$, and $d_{\alpha}$ is the
dimension of the operator $O_{\alpha}$, to be defined precisely later.

\subsection{The effective interactions: Organization of the expansion}

We are interested in models of unconstrained spins. The simplest case
of this type of model is the linear sigma model. This is the model to 
which we explicitly apply all our considerations. 
The canonical surface is 
\be\label{cano_surf}
{\cal H}^0[\phi]=\sum_n \left\{ -\kappa \phi(n) \sum_{\mu} \phi(n+\mu)+
\phi(n)^2+ \lambda ( \phi(n)^2-1)^2 \right \} \ ,
\ee
where $n$ runs all over sites of the lattice, which we take as an infinite 
simple square lattice, and ${\mu}$ runs just over nearest neighbors. 
Spins are unconstrained, i.\ e.\ $\phi(n) \in (-\infty,\infty)$.

Under iteration of RGTs, the new effective interactions ${\cal H}^i$ will 
certainly contain more operators than those three in Eq.~\ref{cano_surf}.
In general, after $i$-RGTs we have
\be\label{ansatz_trans}
{\cal H}^{i}[\vartheta]=\sum_{\alpha} K_{\alpha}^{i} O_{\alpha}(\vartheta) \ ,
\ee
with $\alpha$ running over all possible operators compatible with the 
symmetries of the model. We need an efficient organization of the 
operators appearing in Eq.~\ref{ansatz_trans}.

We consider a basis of operators constructed out of monomials centered at an 
arbitrary site $n$, consisting of powers of fields at site $n$ and products
with fields in other lattice sites. A typical monomial is
\be\label{monom_examp}
{\cal M}_{k,{\cal K}}(n)=\sum_{\rm \{ \vec u,\cdots,\vec w \} \in {\cal K}} 
\phi(n)^{2k}\phi(n+\vec u)\cdots \phi(n+\vec w) \ ,
\ee
where ${\cal K}$ is the class of all vectors that starting
from some representative ones may be constructed out of the symmetries of
the lattice. The full operator is constructed by summing over all lattice 
sites, 
\be\label{oper_examp}
O_{k,{\cal K}}(\phi)=\sum_{n}{\cal M}_{k,{\cal K}}(n) \ .
\ee
We classify all possible monomials (and therefore the operators constructed 
out of them), according to their type, length ($l_{\alpha}$) and  
dimension ($d_{\alpha}$);
\begin{itemize}
\item{\em The type:} We define 4 different types, shown explicitly in 
Table~\ref{tab__deftypes}. Type 0 are bilinear operators. Type
1 are even powers of the lattice field at point $n$. Type 2 and 3 include
products of bilinears and even powers of the field at site $n$.

  \begin{table}[t]
  \centerline{
  \begin{tabular}{|l|c|c|c|}
  \multicolumn{1}{c}{Type 0} &\multicolumn{1}{c}{$\vec u$} &
  \multicolumn{1}{c}{$d_{\alpha}$} & \multicolumn{1}{c}{$l_{\alpha}$}   \\\hline
  $\phi(\vec n)\phi(\vec n+\vec u)$
                            & (1,0)   & 2 & 1  \\
                            & (1,1)   & 2 & 1  \\
                            & (2,0)   & 2 & 2  \\
                            & (2,1)   & 2 & 2  \\
                            & (2,2)   & 2 & 2  \\
                            & .       &   &\\
                            & .       &   &\\
                            & .       &   &\\
                            & $(p,q)$ & 2 & $p$  \\\hline
  \multicolumn{1}{c}{Type 1} &\multicolumn{1}{c}{} &
  \multicolumn{1}{c}{} & \multicolumn{1}{c}{}   \\\hline
   $\phi(\vec n)^{2k}$  &  &  2k  & 0  \\\hline
  \multicolumn{1}{c}{Type 2} &\multicolumn{1}{c}{$\vec u$} &
  \multicolumn{1}{c}{} & \multicolumn{1}{c}{}   \\\hline
  $\phi(\vec n)\phi(n+\vec u)\phi(n)^{2k}$  & (p,q) & 2k+2 & $p$  \\\hline
  \multicolumn{1}{c}{Type 3} &\multicolumn{1}{c}{$\vec u, \vec w $} &
  \multicolumn{1}{c}{} & \multicolumn{1}{c}{}   \\\hline
  $\phi(\vec n+\vec w)\phi(n+\vec u)\phi(n)^{2k}$  & (p,q) , (r,s) & 2k+2 
                    & $   $ \\\hline
  \end{tabular}}
  \caption{ Operators included in the expansion, in the order they
      are considered with their associated dimension and length. The vectors
      $\vec u=(p,q)$ are the representative vectors of the class 
      ${\cal K}$ in Eq.~\ref{monom_examp}. The parameter $k$ starts 
        at 1. }
  \label{tab__deftypes}
  \end{table}

\item{\em The length $l_{\alpha}$: } This is the maximum linear separation 
between fields in a given monomial. An operator with length $l_{\alpha}$ 
can be accomodated in a lattice as small as $(2l_{\alpha}+1)^d$, $d$ being 
the dimensionality of the system.
\item{\em The dimension $d_{\alpha}$: } This is the number of product 
fields in each monomial. 
\end{itemize}

We organize operators within a certain type in increasing length 
and dimension. Overall, we are performing a systematic expansion with the 
assumption that the RG-couplings decay rapidly both with dimension and 
length. We elaborate more on this in considering the properties of RGTs.

There is a last remark concerning the canonical surface Eq.~\ref{cano_surf}.
In this paper, the model will be considered in 2 dimensions. The most 
general RG flow is rather complicated as there are an infinite number of 
inequivalent FPs. From the canonical surface Eq.~\ref{cano_surf},
however, we just can reach two of them, the Gaussian FP (GFP), which is 
the most infrared unstable, and the Ising FP (IFP), which is the most 
infrared stable. To reach any other FP, higher operators, together with a 
fine tuning of the couplings, should be considered.

\subsection{The Bell-Wilson RGT}

There are (infinitely) many different RGTs 
giving rise to different RGTs having different properties.  
The utility of those properties lies in that they may be used to single
out some RGTs as being more accurate than others in approximate calculations 
of the ${\rm T}$ matrix. The choice of a particular RGT is therefore 
determined by the type of approximation we apply.

As we are eventually interested in a finite lattice approximation,
a very important property is to identify RGTs generating very short-ranged 
effective Hamiltonians. That is, at each RG iteration, the effective 
interaction ${\cal H}^i$ generated can be parametrized with a few 
operators of small dimensions and lengths. This is the
case if FP couplings, for example, decrease at least as rapidly as
\be\label{dec_FP_value}
|K_{\alpha}^*| {< \!\!\!\!}_{\sim} 
e^{-d_{\alpha}/\xi_d} e^{-l_{\alpha}/\xi_l} \ ,
\ee
i./e./ exponentially with both dimension and length. We introduced two 
decay corrrelation lengths $\xi_l$ and $\xi_d$. The importance
of this property lies in that the more short-ranged the FP, the more
insensitive it will be to truncation of higher dimension and length
operators. This is the rationale behind the organization of the
expansion of the operators generated along the RG flow in the previous 
section. 

\begin{figure}[t]
\centerline{\epsfig{file=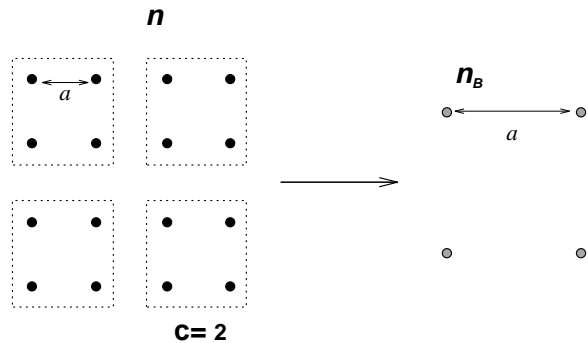,width=3 in}}
\caption{Graphical representation of the short-distance integration
of the RGT Eq.~\ref{Bell_transf}, $n$ labels sites in the fine lattice,
$n_B$ in the coarse one.}
\label{fig__BELL}
\end{figure}

The RGT first introduced in \cite{BELL1} has the property that the
Gaussian FP is short-ranged, 
in the sense of Eq.~\ref{dec_FP_value} \cite{BELL1} \cite{HAS1} 
\cite{THESIS}. It is therefore a natural candidate to investigate 
for other FPs as well. The RGT is
\be\label{Bell_transf}
e^{-{\cal H}^{i+1}[\vartheta]}=\int \prod_{n} d \phi(n) \
e^{-{\cal H}^i[\phi]} 
\ e^{-\frac{a_W}{2}\sum_{n_B}(\vartheta(n_B)-b\sum_{n \in n_B}
\phi(n))^2 } \ ,
\ee
where the average is over $c \times c$ cells in the fine lattice,
with $c=2$ as shown in fig~\ref{fig__BELL}.
There are two a priori free parameters, $a_W$ and $b$. 

The parameter $a_W$ just labels different transformations. Universal
quantities are independent of it. Of course, within an approximation
that may not be the case exactly, and it provides us
with a freedom that we can use to optimize the performance of the RGT.

The parameter $b$, however, has a much deeper meaning related
to the unboundeness of the spins; we may rescale them, and yet the
physics of the model is the same. If there is a FP  we may
generate a whole line of new FPs by rescaling the field $\phi$.
The model shows a redundant direction \cite{WK}, \cite{WEG}.
By definition, any point in that line is a FP of the transformation, 
so the RGT cannot make the FP flow along this direction, and therefore
it is exactly marginal. A first consequence of this is that the ${\rm T}$ 
matrix will always have at least one eigenvalue exactly 
equal to one. The parameter $b$ gets uniquely and self-consistently 
fixed by requiring that there is no flow along this marginal redundant 
direction. Any other choice of $b$ would cause this line to be 
relevant or irrelevant, and there would be no FP in the theory,
in contradiction with the physics of the model.
The rescaling parameter is not a freedom of the model. In fact, it
has a universal meaning. It may be related to the anomalous conformal 
dimension of the field $\eta$, the exponent controlling the
universal algebraic decay as a function of distance of the two-point 
correlation function
\be\label{alg_decay}
G(x)\equiv \langle \phi(x) \phi(0) \rangle \sim \frac{1}{|x|^{d-2+\eta}}
\  \ \ |x| \rightarrow \infty \ ,
\ee
with
\be\label{anom_conform} 
\eta=d+2-2\frac{\ln b^*}{\ln c} \ ,
\ee
where $b^*$ stands for the rescaling factor $b$ at the FP and $c$ will be
usually equal to 2 \cite{WK}.

\section{The RG in a finite lattice}\label{sect__finlat}

\subsection{Theoretical issues}

We consider now the canonical surface Eq.~\ref{cano_surf} and the
transformation Eq.~\ref{Bell_transf} defined on a finite lattice with
periodic boundary conditions (so that translation symmetry is preserved).
The ${\rm T}$-matrix may be computed from Eq.~\ref{der_coup} and
Eq.~\ref{def_matrices}. The question is how much the ${\rm T}$-matrix
computed within this approximation differs from its infinite volume value.

Let us assume that the FP of the RGT is such that the largest 
operator length is $n_s$. The entire FP may be accommodated in a lattice as 
small as $(2n_s+1)^d$. Let us consider it now in a $(2(2n_s+1))^d$ volume,
and apply a RGT, reducing the volume to a $(2n_s+1)^d$. As the 
FP still perfectly fits in, the FP couplings should not differ from their 
infinite volume values. There are no finite size effects \footnote{ More
Rigorously, finite size effects are exponentially suppressed and therefore
very small. They are irrelevant for this argument.}.
This is even more remarkable when one considers that both the couplings and
the ${\rm T}$-matrix are computed from the matrices Eq.~\ref{def_matrices}.
These involve the computation of expectation values that show severe 
finite size effects as the system is critical, and hence has infinite 
correlation length. This may be one of the most interesting
features of MCRG methods. We may, at least in principle, obtain 
infinite volume results working within a finite lattice approximation.
Finite size effects are not dictated by
the correlation length (which is infinite) but by the range of the
FP (and therefore of the RGT). A consequence of the previous arguments
is that the parameters of the canonical surface must be fine tuned 
at criticality. We need to perform the calculations at the values of the 
exact critical surface of the infinite dimensional system, at virtually
infinite correlation length.

Obviously, applying a RGT in a finite lattice reduces its volume. Even if the
RGT is very short-ranged, it may happen that the FP cannot be reached
before the lattice is so small that huge finite-size effects, as explained,
enter into the game. As we may iterate the RGT only a finite number of times,
we need RGTs such that the canonical surface is very close to the FP 
\cite{SWEN2}{\cite{HAS2}.

In short, any RGT to be used in a finite lattice approximation should
fullfill two conditions;
\begin{itemize}
\item {\em Generate short-ranged effective Hamiltonians.}
\item {\em Approach the FP rapidly.}
\end{itemize}

In models of unbounded spins there is still the issue of the determination 
of the rescaling factor $b$ in the RGT Eq.~\ref{Bell_transf}. This
parameter is intimately related to the existence of a line of 
FPs. In general, however, not all FPs along this line will be short-ranged 
enough and the existence of the whole line may be apparent 
only in going to prohibitively large lattices. However, if a
sufficiently short-ranged FP exists, we may expect
at least an infinitesimal portion of this line to survive around it
and the ${\rm T}$ matrix to exhibit at least one eigenvalue equal to one 
with good precision.

\subsection{RG in practice}\label{subsec__prac}

The critical couplings of the canonical surface Eq.~\ref{cano_surf} in
an infinite lattice cannot be computed within a MCRG approach. It is a
quantity to be computed from other techniques, and we extracted it from 
the analysis of existing coefficients of the strong coupling expansion of 
the canonical surface, as discussed in appendix~\ref{app__strong}.

Once the critical couplings are known, iteration of the RGT 
Eq.~\ref{Bell_transf} would drive the canonical surface towards the most 
infrared stable FP, the Ising FP (IFP) in our model. The strategy we 
follow is to pick a value for $a_W$ (see  Eq.~\ref{Bell_transf}) and apply 
as many RGTs as allowed (that is, until the lattice becomes a $4^2$ volume)
 with different values for the rescaling value $b$ (as defined in 
Eq.~\ref{Bell_transf}).  In the previous subsection we discussed the 
different issues that need to be addressed. The different criteria we 
used to tackle them are as follows,

\begin{enumerate} 
\item{\em Short-ranged effective Hamiltonians:} 
\begin{enumerate} \item Using Schwinger-Dyson 
Eq.~\ref{def_coup} we compute the couplings and check for the
ansatz Eq.~\ref{dec_FP_value}. 
\end{enumerate}

\item{\em Rapid approach to the FP:} There are two different criteria to
be met.
	\begin{enumerate} 
             \item If a FP is reached, the couplings of these effective 
               Hamiltonians after successive applications of RGTs
	       should coincide.
            
	     \item
              We perform RGTs starting from the same canonical surface
              but in a smaller lattice and compare expectation values
              of operators at the same volume size. If a FP is reached 
              these expectation values should agree.
        \end{enumerate}

\item{\em The determination of the anomalous dimension $\eta$}
	\begin{enumerate} 
	    \item
	    The ${\rm T}$ matrix should exhibit an
            exactly marginal eigenvalue at the right value of the rescaling
            factor.
            \item
            Similarly as in criterion 2(b), the expectation values of
            operators should agree at the correct value of
            the anomalous dimension.

          \end{enumerate} 
\end{enumerate}

We then scan over different values of $a_W$ and select the values
optimally fulfilling all previous criteria.

Criteria 1(a) and 2(a) are obvious. Criteria 2(b) uses the fact that
although finite size effects for these expectation values are huge,
they enter in exactly the same functional form \cite{SHEN1}, and if
a FP is reached, those expectation values should agreee.
Criterion 3(a) to pick up the anomalous dimension is a modification 
of \cite{BELL1}. It was first introduced in \cite{LAT97} and successfully 
applied to the linear sigma model in 3 dimensions, although the 
lattices used were slightly too small, and finite size effect errors 
were difficult to assess. Criterion 3(b) is a necessary consequence of the 
discussion in criterion 2(b) and the fact that a wrong choice for $b$ 
should not lead to a FP at all. 

Besides the statistical errors inherent in any numerical simulation,
there are different sources of putative systematic
errors,

\begin{itemize}
\item{\em e(1) Truncation errors:} If some operator having a sizeable
coupling is not included in Eq.~\ref{der_coup} and Eq.~\ref{def_coup}
the ${\rm H}$ and ${\rm M}$ matrices miss sizeable matrix elements
which translates into systematic errors in the determination of both
the ${\rm T}$ matrix and the couplings.

\item{\em e(2) Finite size effects:} This error appears when 
the lattice is too small to accommodate operators having sizeable 
couplings. 

\item{\em e(3) Off-criticality:} If the couplings in the starting
Hamiltonian are not tuned to criticality, the flow eventually converges 
towards the renormalized trajectory, moving away from the FP.

\item{\em e(4) Lack of eigenvector convergence:} 
It may happen that some eigenvectors of the ${\rm T}$ matrix are just not 
convergent, in the sense that the coefficients defining it have just a
formal meaning. This may translate into the correponding eigenvalue being 
completely unreliable.

\item{\em e(5) Redundant operators:} Some eigenvalues may depend on the
transformation, the associated critical exponent is then just an artifact
of the transformation \cite{SHAN1},\cite{GOCK1}.

\end{itemize}

Finally, let us recall that in models of unconstrained spins, like the ones 
we are interested in, the rescaling itself will have an associated error bar  
due to statistical errors and e(1),e(2) and most importantly e(3). This
rescaling error bar the will translate into an additional 
error in computing other quantities.

\section{Numerical details}\label{sect__numdet}

\subsection{The algorithm}

We use an embedding algorithm \cite{BROW1}, each step consisting of
15 cluster-Wolff updates \cite{WOLF1} and 5 metropolis hits. We need
20 sweeps of this type to completely decorrelate data.  
We computed the error as a function of bin-size using
standard Jackknife techniques and found a 
flat plateau showing that data in different bins are totally uncorrelated. 
After thermalization,  we generated 64,000 decorrelated  
$64^2$ lattice configurations and 50,000 $34^2$ lattice configurations 
at criticality, and stored them in disk.  The uncompressed storage size of 
these is about 2GB and 500MB respectively.
The simulation was carried out on a 233 MHz Pentium processor, and took
about 2 days of total CPU time.

The RGT Eq.~\ref{Bell_transf} is implemented using the shift
\be\label{RG_noise}
\vartheta(n_B)=b\sum_{n \in n_B}\phi(n)+\frac{\zeta}{\sqrt{a_W}} \ ,
\ee
where $\zeta$ is a randomly distributed gaussian variable with zero mean
and variance 1. From Eq.~\ref{RG_noise} we generate the successive
RGTs. On a 233 MHz Pentium processor it takes about 100 minutes of 
CPU time to generate the ${\rm M}$ matrices
and ${\rm H}$ matrices including up to 40 operators 
for $32^2$, $16^2$, $8^2$ and $4^2$ volumes.

\subsection{The method}\label{sect__method_num}

The canonical surface Eq.~\ref{cano_surf} is a actually a line
$\kappa(\lambda)$. We performed our simulations at
$\lambda=1.0$, with corresponding  $\kappa(1.0)=0.6795$.

Concerning the expansion of the effective Hamiltonians generated along the
RG flow, we considered up to 40 operators in total. All operators of type 0 
up to length 3,  operators of type 1 up to dimension 10, operators 
of type 2 up to length 3 and dimension 6, 
and operators of type 3 up to 
length 2 and dimension 4, 
as defined in Table \ref{tab__deftypes}. In the
discussion that follows we will identify these operators by number from 
1 to 40.  Operators 1 to 5 correspond to type 1 $k= 1,...5$.
Operators 6 to 14 correspond to type 0 operators (p,q)$=(1,0),..,(3,3)$.
Operators 15 to 23 correspond to type 2 operators with $k=1$ 
(p,q)$=(1,0),..,(3,3)$.
Operators 24 to 32 correspond to type 2 operators with $k=2$ 
(p,q)$=(1,0),..,(3,3)$. Operators 33 to 40 are the eight distinct $k=1$
type 3  operators that fit in a $3 \times 3$ section of lattice 
($\vec n$ at the center): [(p,q)(r,s)] = [(0,1)(0,1)], [(0,1)(1,0)],
 [(0,1)(0,-1)], [(0,1)(1,1)], [(0,-1)(1,1)], [(1,1)(-1,1)], and 
[(1,1)(-1,-1)].

We systematically explored anomalous 
dimensions $\eta=0.10$, $0.20$, $0.25$, $0.30$, $0.40$, $0.50$. 
In some particular cases additional values were also considered. 
To improve the performance of the transformation, we studied the 
following values of the RGT parameter:
 $a_W=8,16,20,40,80,\infty$.

In order to efficiently handle the enormous amount of data generated, 
we organized the information in different {\em html}-tables, each one
displaying the particular property under study. Samples of tables are
available at the address 
\quad http://web.syr.edu/\~{\hspace{-.05cm}}acacciut

\section{Results}\label{sect__results}

In this section we present an extensive and detailed analysis of the 
methods and techniques already introduced. We keep the discussion 
technical. The main results may appear a little dispersed, so in
the next section we provide a presentation of the
most relevant final results and the different implications they 
may have.

\subsection{Short-ranged Hamiltonians}

The effective interactions generated along the RG flow depend on
$a_W$ (the transformation) and $b$ (the rescaling). The subject under
study now is the decay of couplings with dimension and length as
a function of $a_W$ for a given $b$. The optimal RGT will be the 
one generating the most short-ranged effective interaction.

The first issue to address is the dependence of couplings on length.
All the different values of $a_W$ examined conform to the decay ansatz
Eq.~\ref{dec_FP_value}, although the decay is too fast to allow a
precise determination of the correlation length decay $\xi_l$, as coupling
constants are zero within statistical error bars for 
$l \geq 2$. The typical situation ocurring for any value of 
$a_W$ is plotted in \footnote{We generalize slightly
the definition of length by giving length $\sqrt{2}$ to the operator $(1,1)$
instead of $1$, as defined.} fig.~\ref{fig__type0dis}.
From the data in fig.~\ref{fig__type0dis} we can give a very conservative 
upper bound for the correlation length decay, 
\be\label{lower_xil}
  \xi_l <\frac{1}{2} \ ,
\ee
valid for any value of $a_W$. We conclude that operators have a 
very good decay property for any $a_W$ value as far as length 
is concerned. 

\begin{figure}[t]
\centerline{\epsfig{file=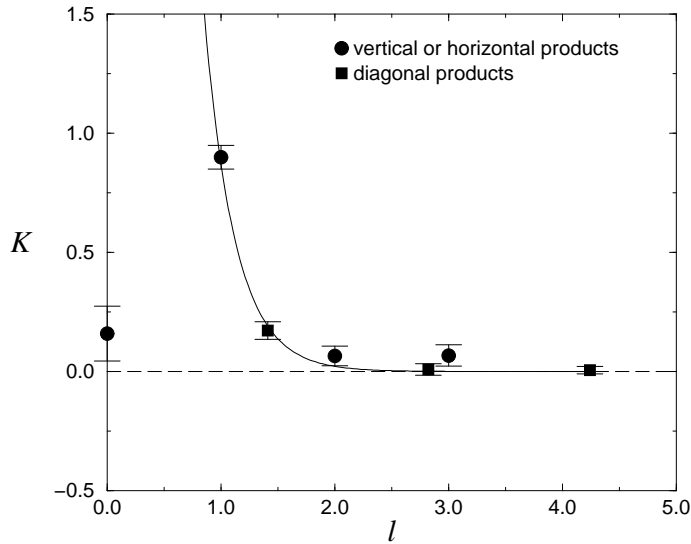,width=4 in}}
\caption{Plot for $a_W=16$, $\eta=.30$ of the magnitude of type 0 
(bilinear) couplings as a function of length for 3 RGTs. Squares are 
vertical or horizontal products and circles the diagonal ones.
The solid line is an exponential decay.}
\label{fig__type0dis}
\end{figure}

Couplings as a function of dimension show a very strong 
dependence on $a_W$. In fig.~\ref{fig__type1dis} a logarithmic
plot of the magnitude of type 1 couplings is shown as a function of 
dimension. For small $a_W$ the decay ansatz, Eq.~\ref{dec_FP_value}, sets 
in for small values of dimension. For $a_W=8$, we extract a dimension 
correlation length decay
\be\label{xi_dimensions}
\xi_d \sim 1.0 \ .
\ee
The correlation length decay as a function of distance is slower than
with length.
Increasing $a_W$ further, there is a transient increase of couplings, 
with a peak moving forward in dimension as a function of $a_W$, before
the asymptotic form Eq.~\ref{dec_FP_value} becomes apparent, as shown
in fig.~\ref{fig__type1dis}. In particular, the RGT at $a_W=\infty$ shows
a remarkably poor behavior.

\begin{figure}[t]
\centerline{\epsfig{file=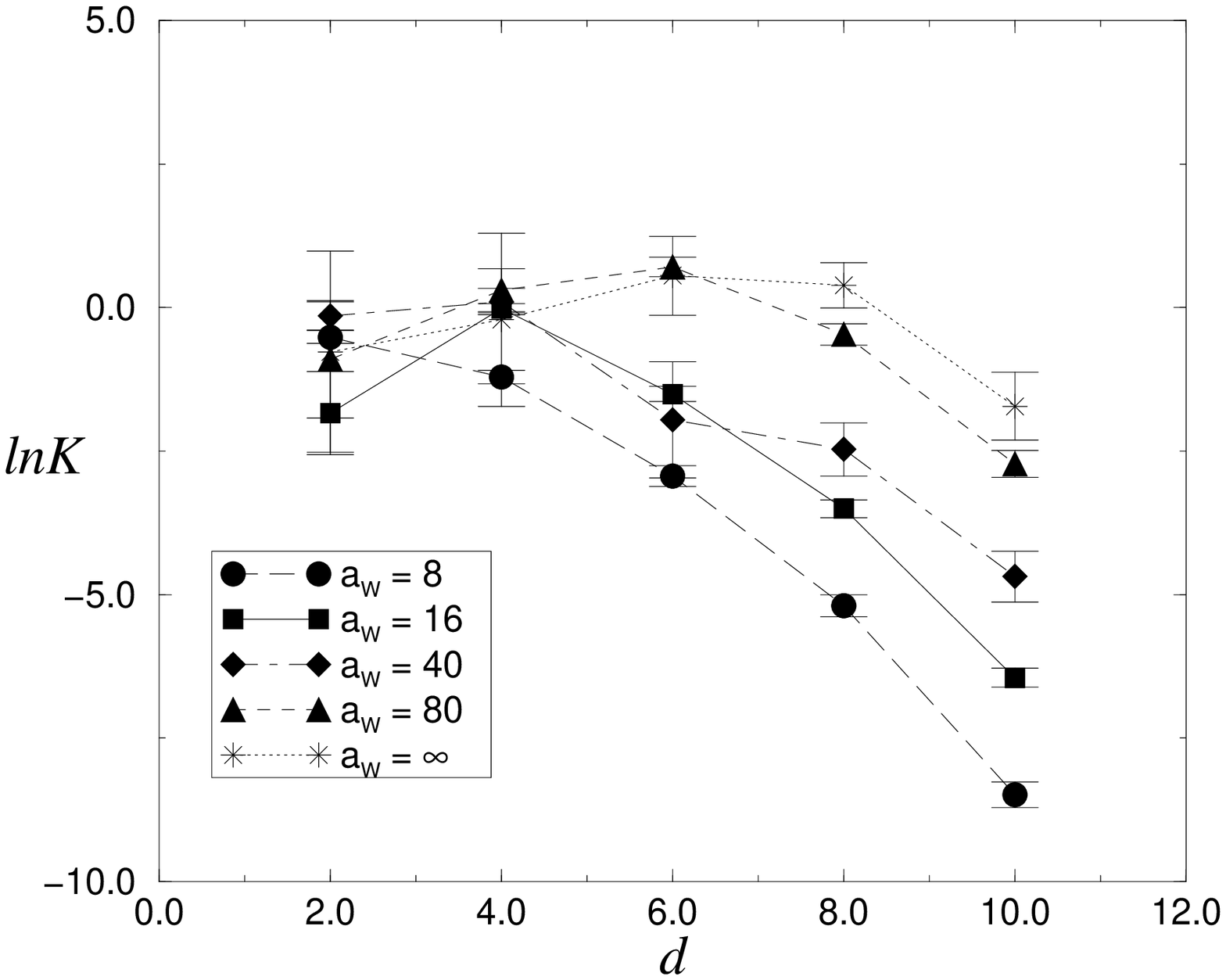,width=4 in}}
\caption{The logarithmic plot of the magnitude of type 1 couplings against 
dimension, $d$.  Results are for 3rd RGT at $\eta=.30$.}
\label{fig__type1dis}
\end{figure}

We cross-check previous results by studying the decay of
type 2 operators. As shown in fig.~\ref{fig__type2}, 
couplings decay exponentially, with a correlation length completely
compatible with the bound in Eq.~\ref{lower_xil}.

\begin{figure}[t]
\centerline{\epsfig{file=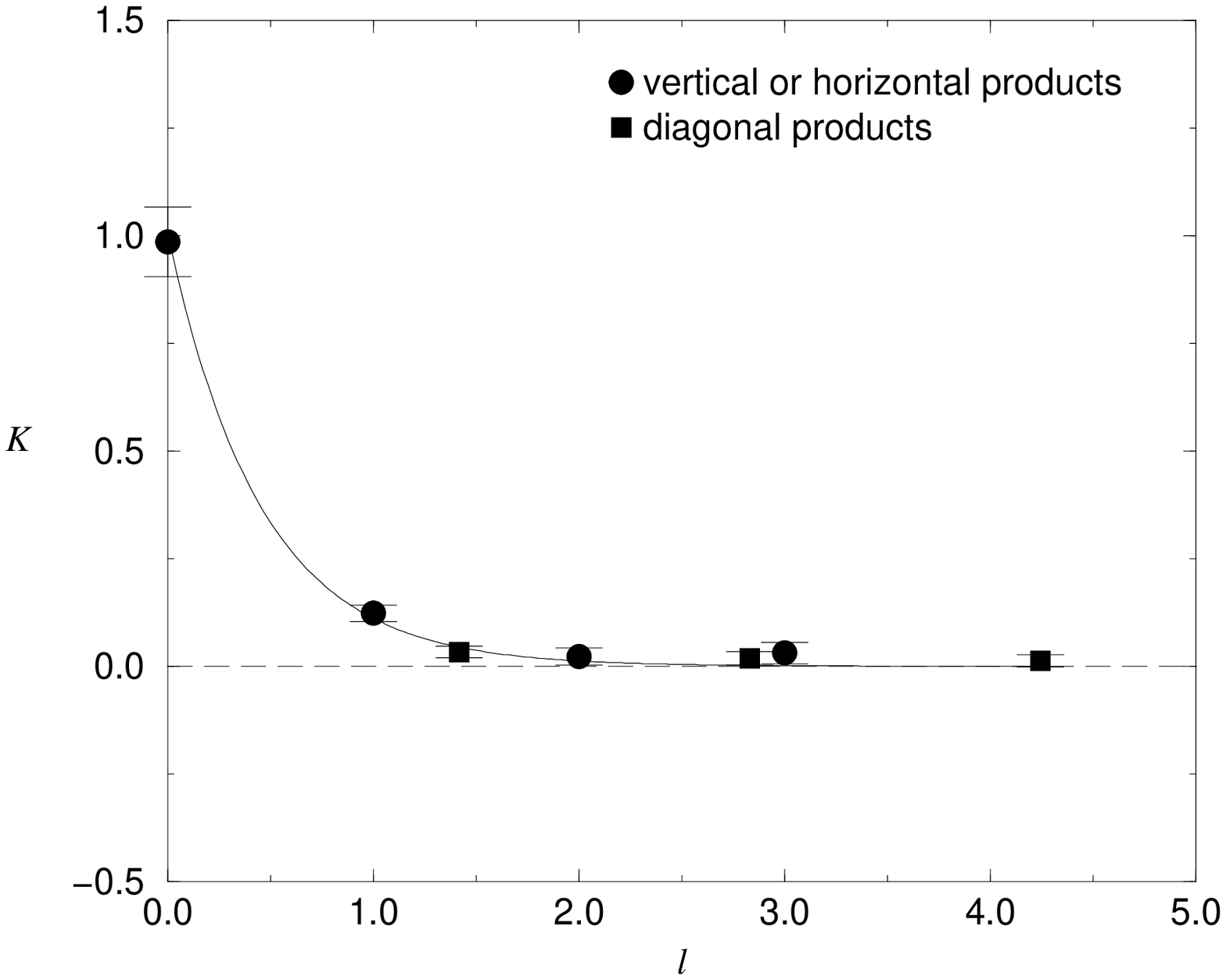,width=2.8 in,
}\epsfig{file=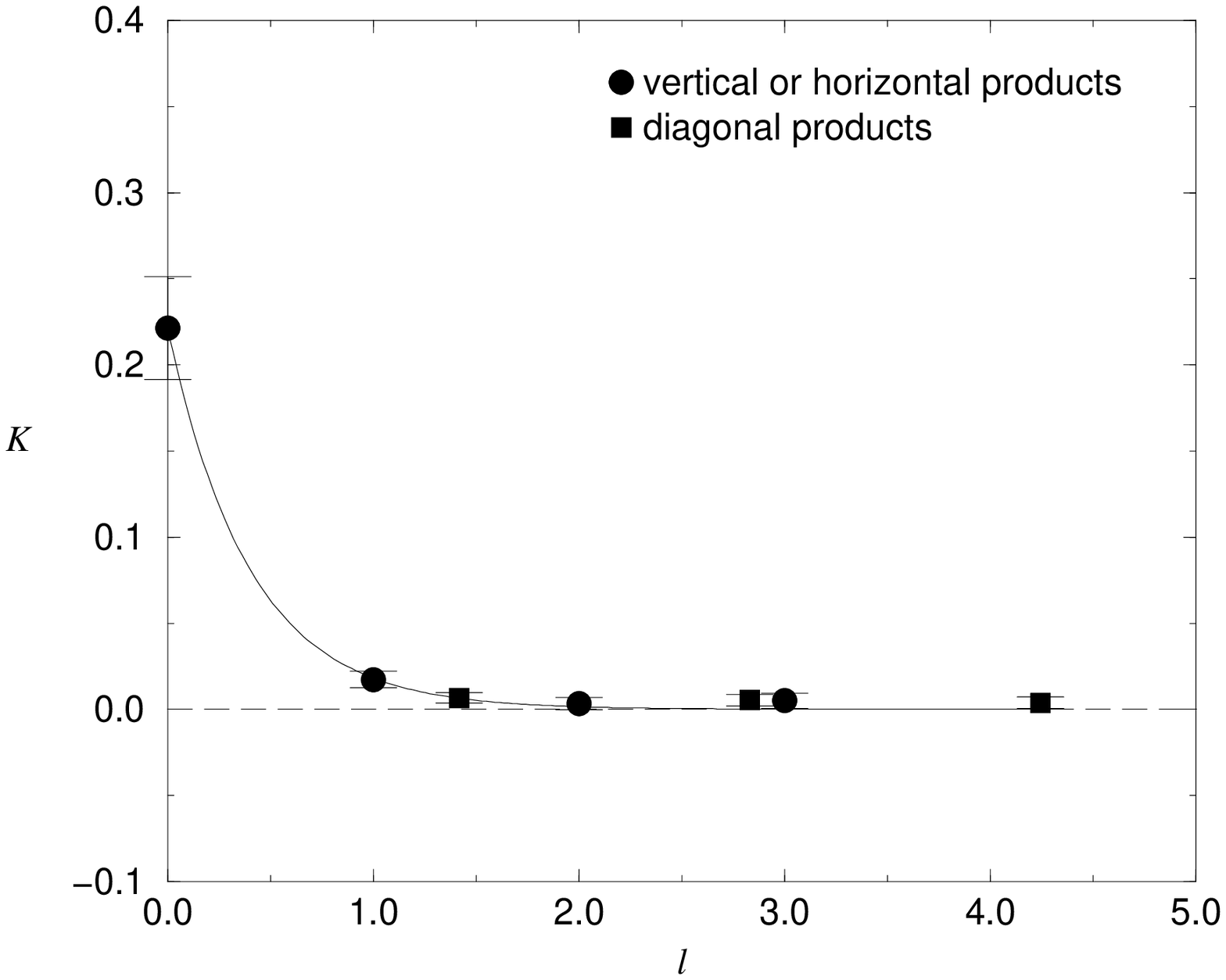, width=2.8 in }}
\caption{Decay of type 2 couplings against length for $k=1$(left) and
$k=2$(right) (see table \ref{tab__deftypes} for definitions)
operators against length. Results are for 3 RGTs at $a_W=16$ $\eta=.30$.
The solid line is the exponential fit, which is in agreement with 
Eq.~\ref{lower_xil}.}
\label{fig__type2}
\end{figure}

From all the evidence accumulated, it is clear that smaller values 
$a_W$ restrict the dependency of the effective interactions to lower 
dimension field operators.
In particular, effective interactions generated with
RG transformation with $a_W \geq 80$ are not particularly 
recommended, as the couplings decay too slowly with dimension to obtain 
reliable results.

\subsection{Fast approach to the FP}

The effective interactions show a double dependence on the rescaling
$b$ and on the parameter $a_W$. The dependence on $b$ is
related to the determination of the anomalous dimension and will
therefore be studied in the next subsection. Now we content 
ourselves with identifying the RGTs that converge rapidly to putative 
FPs.

Concerning criterion 2(a), the matching between couplings in 
successive RGTs, although the amount of information that may be
extracted is certainly limited by the somewhat large statistical errors
in the couplings, it is still powerful enough to provide strong
evidence that we approached a FP after a few RGTs.

For values of $a_W \geq 80$ couplings from successive RGTs do not 
match satisfactorily, and we cannot identify a
FP for any value of the rescaling. In agreement with previous results, 
RGTs with large $a_W$ values perform very poorly. We move on to
investigate lower values for $a_W$. Results at $a_W=40$ do not
show a clear FP for any value of the rescaling, as there are some
couplings that do not agree within 2$\sigma$(statistical), see 
table~\ref{tab__aw_40}. 

\begin{table}[htb]
\centerline{
\begin{tabular}{|c||l|l|l|l|}
\multicolumn{1}{c}{$\eta$}   &
\multicolumn{1}{c}{$1$}   & \multicolumn{1}{c}{$9$} &
\multicolumn{1}{c}{$11$} & \multicolumn{1}{c}{$18$} \\\hline
0.25  & $-0.6(2)   $ & $0.02(5)$  & $-0.03(7)$  & $0.002(20)$  \\
      & $-1.1(3)   $ & $0.19(4)$  & $0.15(8) $  & $-0.08(3) $  \\\hline
0.30  & $-0.40(20) $ & $-0.03(7)$ & $-0.03(7)$  & $-0.005(20)$ \\
      & $-0.87(21) $ & $0.12(4)$  & $0.14(8) $  & $-0.09(2)$    \\\hline
0.40  & $-0.40(20) $ & $0.03(5)$  & $-0.03(6)$  & $-0.003(20)$  \\
      & $-0.56(24) $ & $0.17(3)$  & $0.12(7) $  & $-0.06(2)$  \\\hline
\end{tabular}}
\caption{Example of couplings at 2(up) and 3 RGTs(down) for $a_W=40$ that 
         match poorly.}
\label{tab__aw_40}
\end{table}

For $a_W=8$, there is no clear evidence for a FP at $\eta \geq .30$, see 
table~\ref{tab__aw_20}, although we find an acceptable matching 
for smaller values of $\eta$. 

\begin{table}[htb]
\centerline{
\begin{tabular}{|c||l|l|l|l|}
\multicolumn{1}{c}{$\eta$}   &
\multicolumn{1}{c}{$1$}   & \multicolumn{1}{c}{$18$} &
\multicolumn{1}{c}{$21$} & \multicolumn{1}{c}{$30$} \\\hline
0.30  & $0.67(7)   $ & $0.014(6)  $ & $0.001(5) $  & $0.0001(5)$  \\
      & $0.60(6)   $ & $0.00(1)   $ & $0.015(6) $  & $-0.0020(7) $  \\\hline
0.40  & $0.52(6)   $ & $0.012(5)  $ & $-0.001(4)$  & $-0.0000(4)$ \\
      & $0.38(6)   $ & $-0.003(4) $ & $0.012(4) $  & $-0.0015(6)$    \\\hline
0.50  & $0.39(6)   $ & $0.010(4)  $ & $0.001(4) $  & $-0.0000(4)$  \\
      & $0.19(6)   $ & $-0.003(3) $ & $0.010(4) $  & $-0.0010(4)$  \\\hline
\end{tabular}}
\caption{Example of couplings at 2(up) and 3 RGTs(down) for $a_W=8$ with 
         rather poor matching.}
\label{tab__aw_20}
\end{table}

The value of $a_W=16$ shows putative FPs within the statistical errors 
for values of the anomalous dimension in the range $\eta=.20-.40$.
As an example, in fig.~\ref{fig__eta25__aw16match} couplings from 2 and
3 RGTs are plotted at $\eta=.25$. It is apparent that most of the couplings
agree very well. It seems then that $a_W=16$ is an optimal choice according 
to criterion 2(a).

\begin{figure}[htb]
\centerline{\epsfig{file=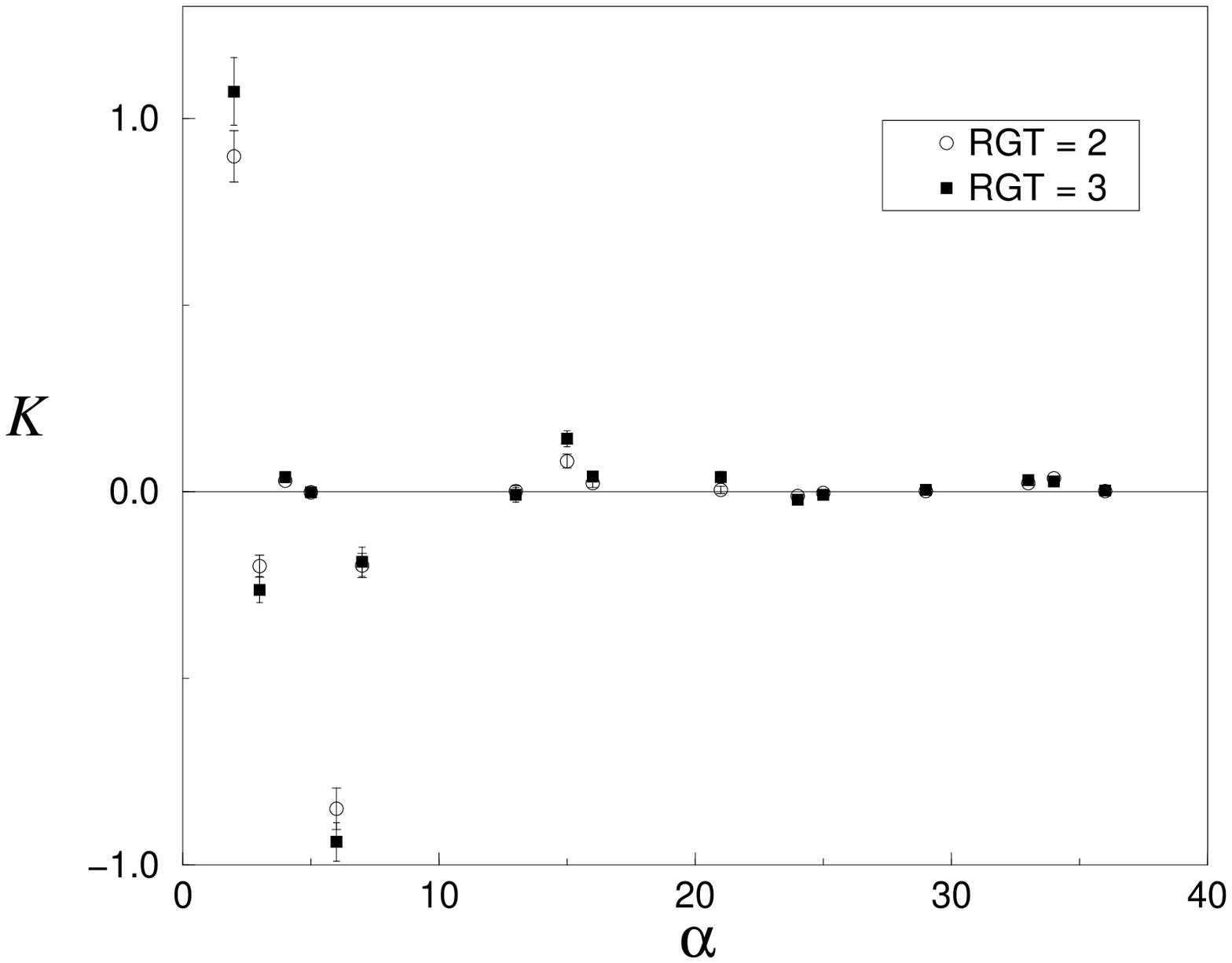,
width=\textwidth }}
\caption{Some chosen couplings at $a_W=16$ and $\eta=.25$ for 2 and 3
RGTs.}
\label{fig__eta25__aw16match}
\end{figure}

Concerning criterion 2(b), the matching of expectation values of 
operators computed using different lattice sizes, we start by defining 
the quantitities
\bea\label{Quality_factor}
r_{\alpha}^{(i)}&=&\left|\frac{\langle O_{\alpha}^{(i)} \rangle-
\langle O_{\alpha}^{(i-1)} \rangle }
{\langle O_{\alpha}^{i}\rangle}\right| \ , \ 
\nonumber\\
Q^{(i)}&=&\sum_{\alpha}(r_{\alpha}^{i})^2 \ ,
\eea
which measure the quality of the matching for expectation values of
operators. A first insight is gained from the plot of the factor $Q^{(i)}$,
(see fig.~\ref{fig__sigma_match_all}). There is a clear minimum  
independent of $a_W$ for $\eta \sim 0.3$. This matching procedure 
appears to be a very precise criterion to fix the rescaling, but this 
a question to be considered later.
What is more surprising from fig.~\ref{fig__sigma_match_all} 
is the very good matching of observables for any value of $a_W$
including values for which we do not find evidence for a FP from
criterion 2(a). However, criterion 2(b) is completely insensitive to
the truncation of operators and to finite size effects, the systematic 
errors that affect criterion 2(a) the most, but
is very sensitive to the canonical surface not being critical, which
probably accounts for the small systematic error in $\eta$ (the exact
value is $\eta=0.25$). A closer inspection, reveals that the quality 
of the matching is sensitive to $a_W$. As apparent from 
fig.~\ref{fig__aw_sensitivity}, a careful comparison shows that the approach 
to the FP is faster for a transformation at $a_W=16$.

\begin{figure}[p]
\centerline{\epsfig{file=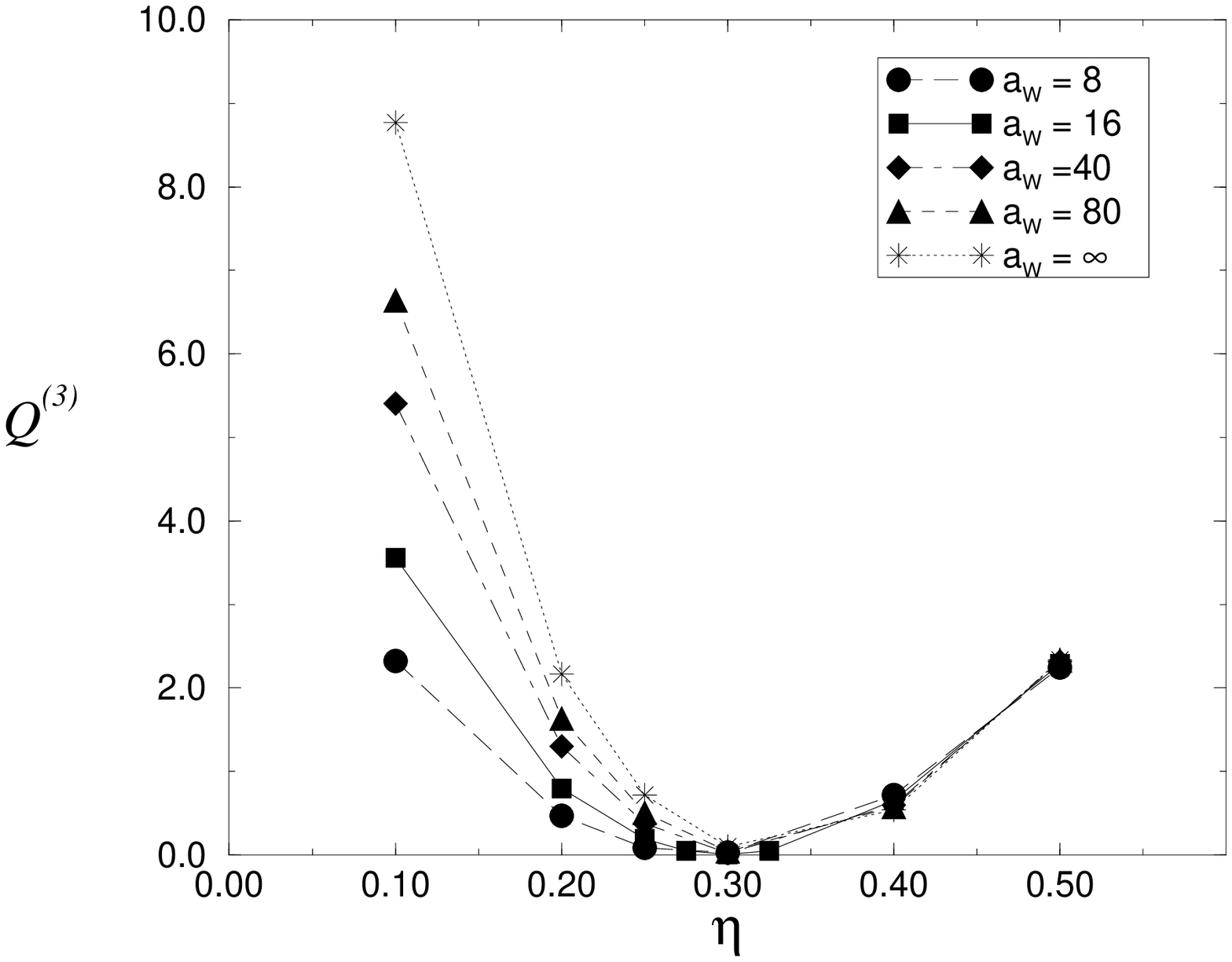,width=3.6 in}}
\caption{Dispersion in the matching of expectation values of 
operators, as compared from 3RGTs starting at $64^2$ and 2 RGTs starting
at $32^2$.}
\label{fig__sigma_match_all}

\centerline{\epsfig{file=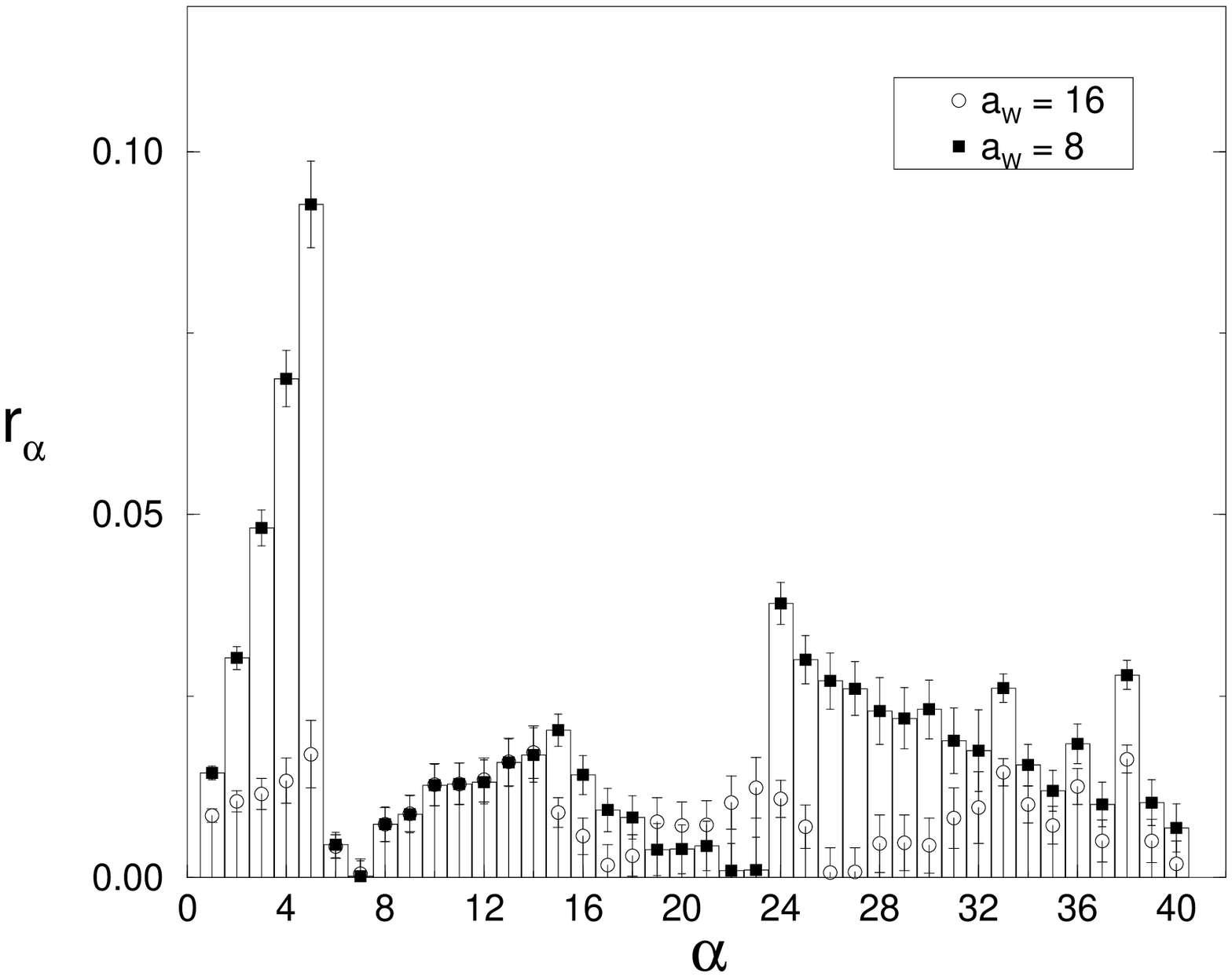,width=3.6 in}}
\caption{Comparison of the quantity $r_{\alpha}$, as defined in  
Eq.~\ref{Quality_factor} for $a_W=8$ and $a_W=16$ 
as compared from 3RGTs starting at $64^2$ and 2 RGTs starting
at $32^2$. The RGT with $a_W=16$ produces a better matching.}
\label{fig__aw_sensitivity}
\end{figure}

To cross-check criterion 2(b) further, we compared the matching of 
observables on the next RGT, that is, observables computed in a $4^2$
lattice. The number of operators available gets reduced, 
as some of them can no longer be accommodated in such small lattice,
but for those that it exists, the matching should significantly improve
as compared with the previous RGT. This is, indeed, corroborated
from fig.~\ref{fig__perf__match}, the matching in this case is 
close to perfect.

\begin{figure}[hp]
\centerline{\epsfig{file=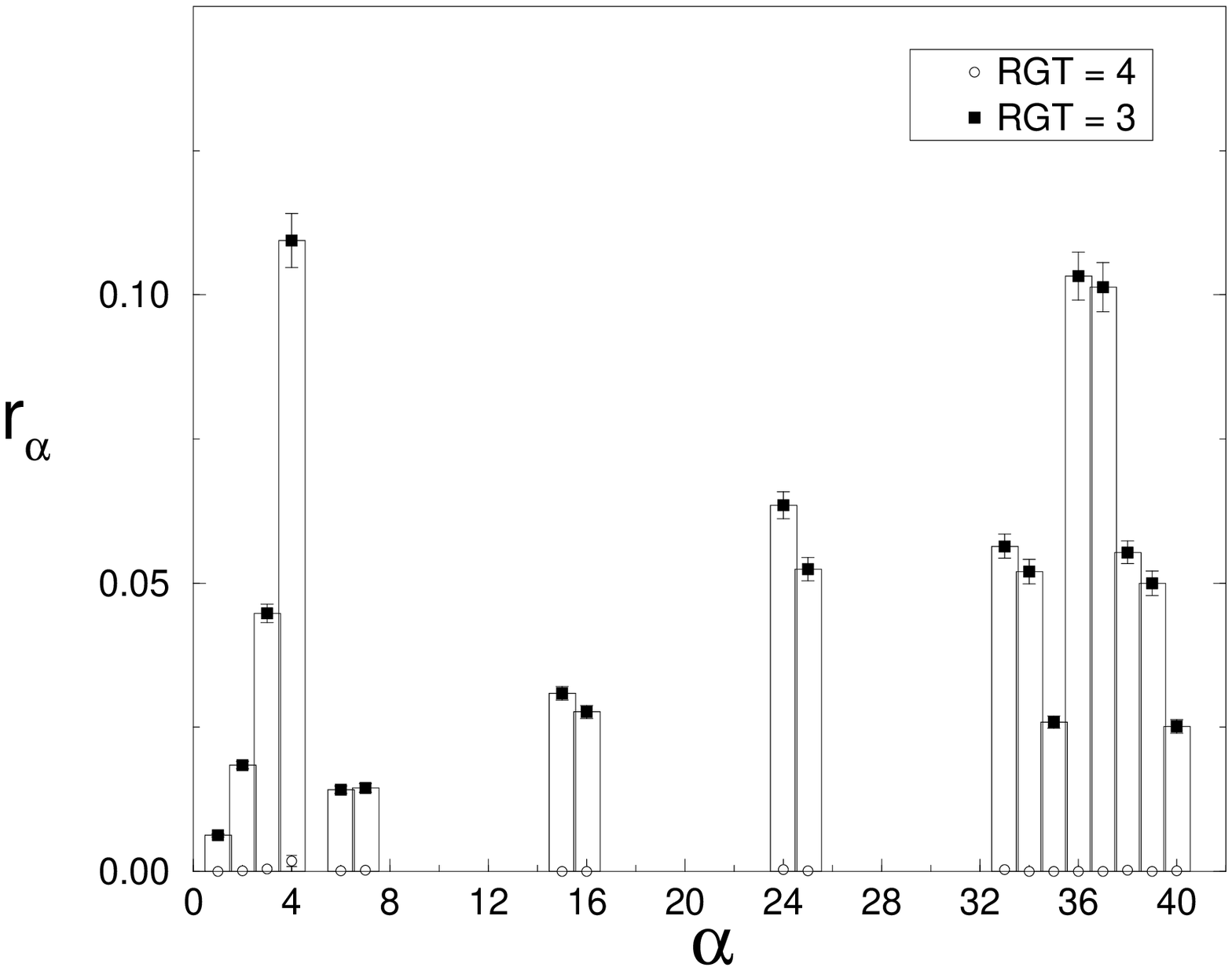,
width=\textwidth }}
\caption{Comparison of $r_{\alpha}$ after 4/3($4^2$ lattice)
and 3/2 RGT ($8^2$) lattice at $\eta=0.30$. Only operators 
that can be accommodated in a $4^2$ lattice are plotted. 
The matching is nearly perfect.}
\label{fig__perf__match}
\end{figure}

\subsection{The determination of the anomalous dimension}

Previous considerations have selected the value of $a_W=16$
as the one best compromising the 2 criterion of short-rangeness and
fast approach to the FP. We now come to the determination of the
rescaling factor, which as a byproduct allows us to compute the
first critical exponent of the theory, the conformal anomalous 
dimension, see Eq.~\ref{anom_conform}. We 
devised two different criterion to be meet.

The first criterion 3(a) is the proper identification of a marginal
eigenvalue in the ${\rm T}$ matrix. In our particular problem, this
is complicated by the fact that in the cases were the two criteria
we use to determine that a FP is reached,
we clearly get two marginal directions. It seems natural to assume generally 
that the model has actually two marginal directions. 
Although we already ruled out some values 
of $a_W$, we keep the discussion general for the time 
being and consider all values. The second eigenvalue of the 
{\rm T} matrix as a function of $\eta$ is plotted in 
fig.~\ref{fig__sec__fun} for 2 RGTs. Large values of $a_W$ are 
incompatible with the ${\rm T}$ matrix having a marginal eigenvalue, except 
for large values of the anomalous dimension, where the third eigenvalue is
far from one, and following previous discussions should be disregarded. 
This is reassuring, since for those values of $a_W$ we already know that 
RGT perform rather poorly.
For smaller values of $a_W$, we do get a one eigenvalue
within statistical error bars in the range $.10 < \eta < .40$. 
If a FP is reached, the marginal eigenvalue should be stable through the
next RGTs, plotted in fig.~\ref{fig__third__fun}. Also, for small
values of $\eta$, which criterion 3(b) will rule out as well, the
third eigenvalue is clearly different from one.  Within statitical
error bars, the window of anomalous dimensions is further restricted
to be in the range $0.25\leq \eta \leq 0.30$. 

\begin{figure}[p]
\centerline{\epsfig{file=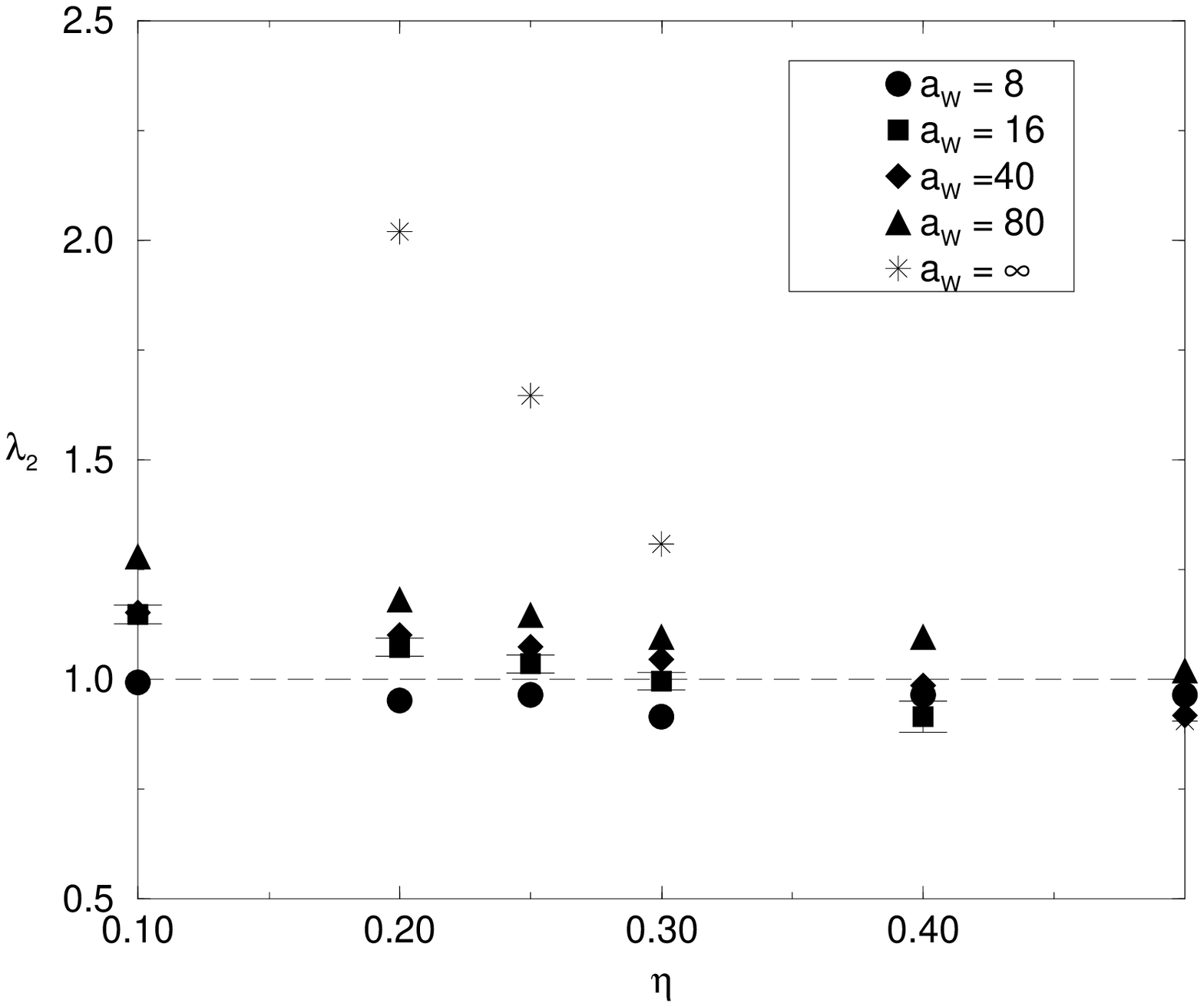, width=3.6 in}}
\caption{The second eigenvalue of the ${\rm T}$ matrix as a function of
the anomalous dimensions for different values of $a_W$ after 2 RGTs.}
\label{fig__sec__fun}

\centerline{\epsfig{file=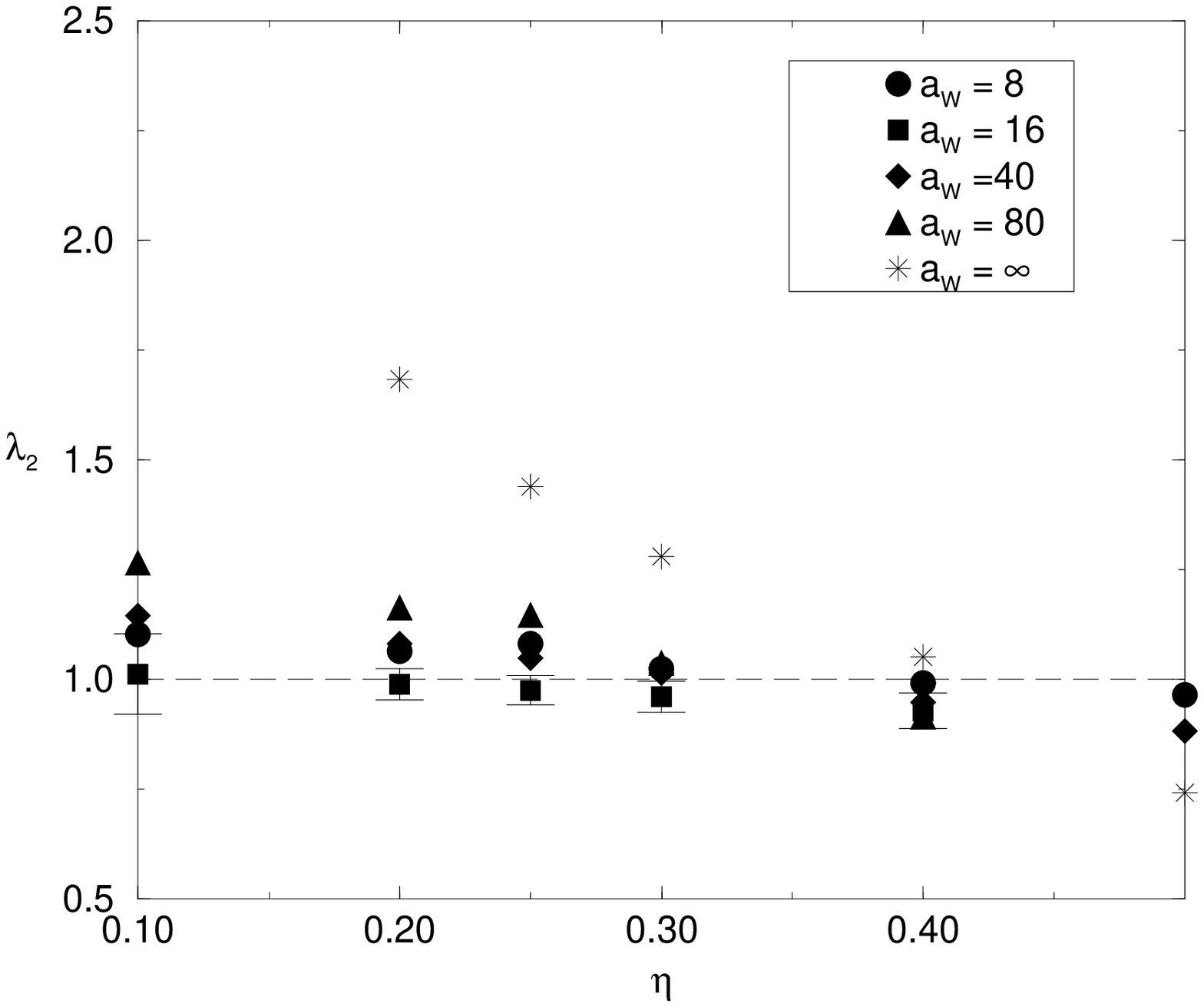,width=3.6 in}}
\caption{The second eigenvalue of the ${\rm T}$ matrix as a function of
the anomalous dimensions for different values of $a_W$ after 3 RGTs.}
\label{fig__third__fun}
\end{figure}

The matching between operators, criterion 3(b) turns out to be the 
most sensitive criterion, a result already anticipated in
fig.~\ref{fig__sigma_match_all}. For our favorite value
$a_W=16$, all expectation values of operators match for $\eta=.30$ 
within 2$\sigma$ in the statistical error bars,
for all 40 operators considered, see
fig.~\ref{fig__match_two}. For comparison, we plot as well the
relative error for values of the anomalous dimensions $\eta=0.10$
and $\eta=0.50$ in which there is no matching at all.
Let us recall that if we perform another RGT, the matching is perfect
see previous fig.~\ref{fig__perf__match}. This provides an important
cross-check that we are in the vicinity of the IFP. Unfortunately, the
final lattice ($4^2$) is too small and shows finite size effects in
the critical exponents, as we will discuss.

\begin{figure}[p]
\centerline{\epsfig{file=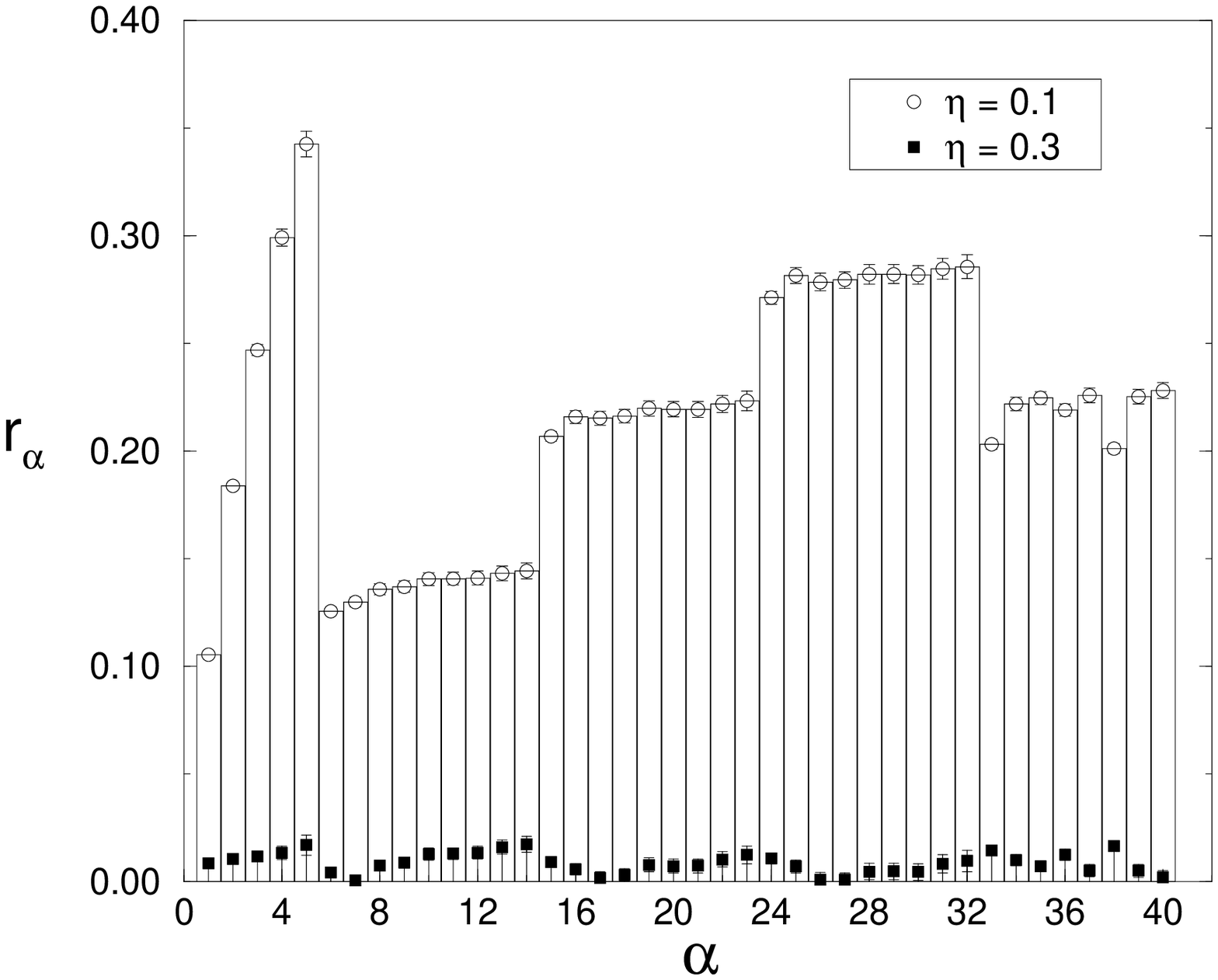,width=3.7 in}}
\centerline{\epsfig{file=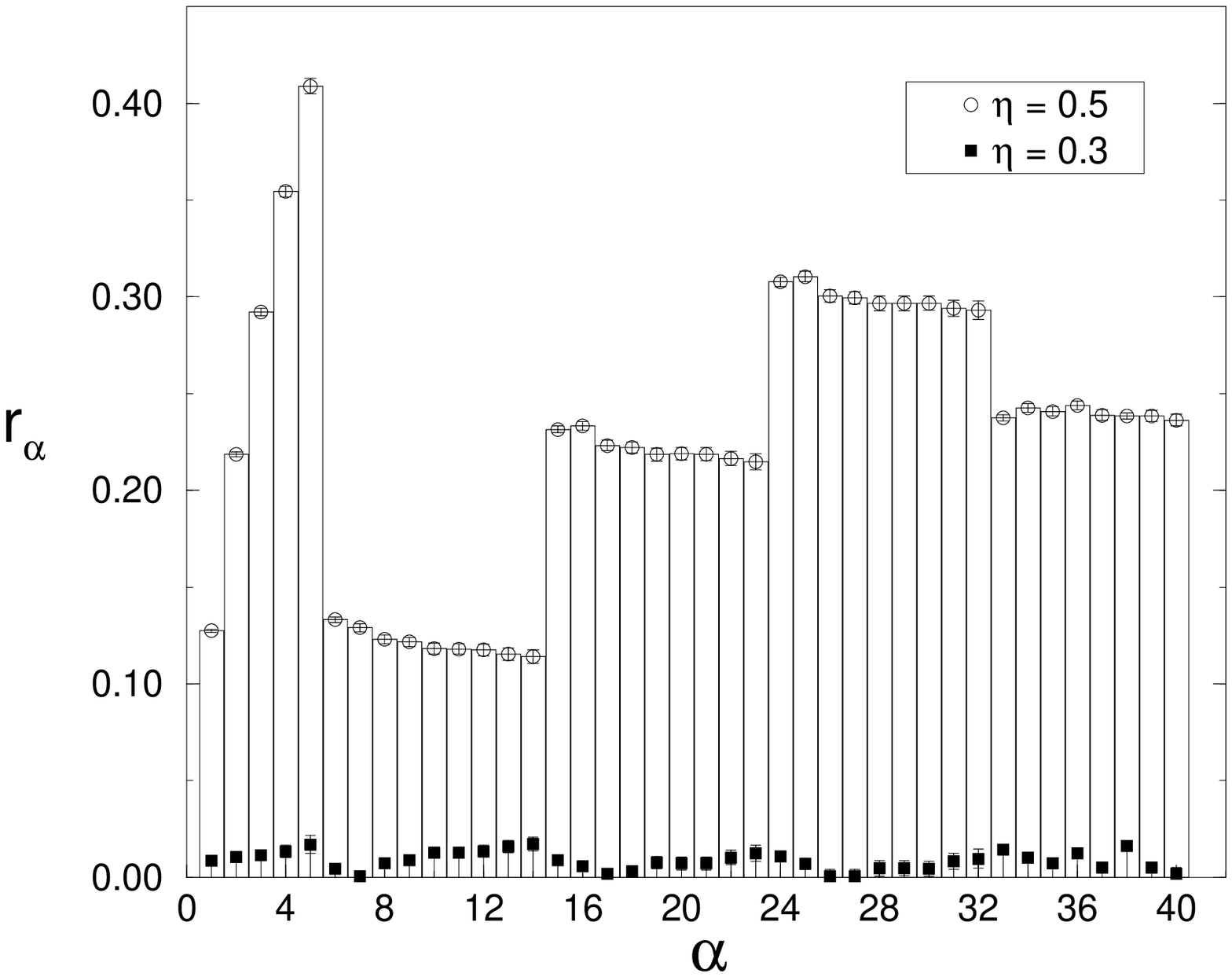, width=3.7 in}}
\caption{Plot of the quantity $r_{\alpha}$ as defined in 
Eq.~\ref{Quality_factor} comparing the excellent matching at $\eta=0.30$
with no matching at all at $\eta=0.10$(top) and $\eta=.50$(bottom).
The final lattice is $8^2$.}
\label{fig__match_two}
\end{figure}

The matching criterion  singles $\eta \sim 0.30$ as the preferred
value for the anomalous dimension, a result slightly off from
the exact value, $\eta=.25$. The matching at $\eta=.25$ is also good 
but not as good as in $\eta=.30$. We will expand further on this issue
in the discussion of the different sources of systematic errors.

\subsection{The critical exponents and the FP Hamiltonian}

The rescaling parameter has been computed and it
corresponds to an anomalous dimension in the interval 
$\eta \in (.25,.30)$, our results favoring values closer to
$\eta=0.30$ than to $\eta=.25$. 

\begin{figure}[hp]
\centerline{\epsfig{file=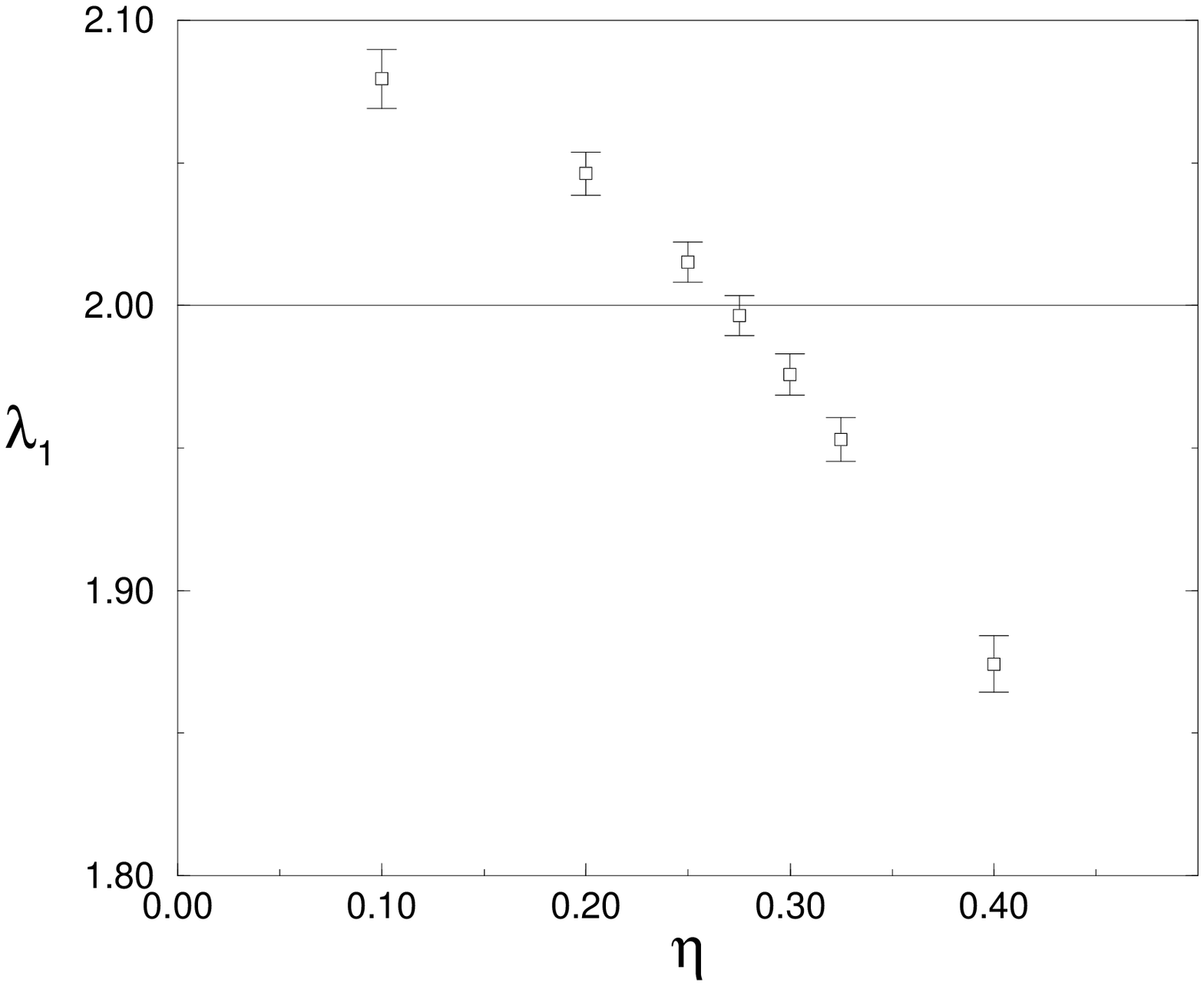,width=3.6 in}}
\caption{First eigenvalue as a function of the anomalous dimension 
at $a_W=16$, final lattice is $8^2$.}
\label{fig__first_eta}

\centerline{\epsfig{file=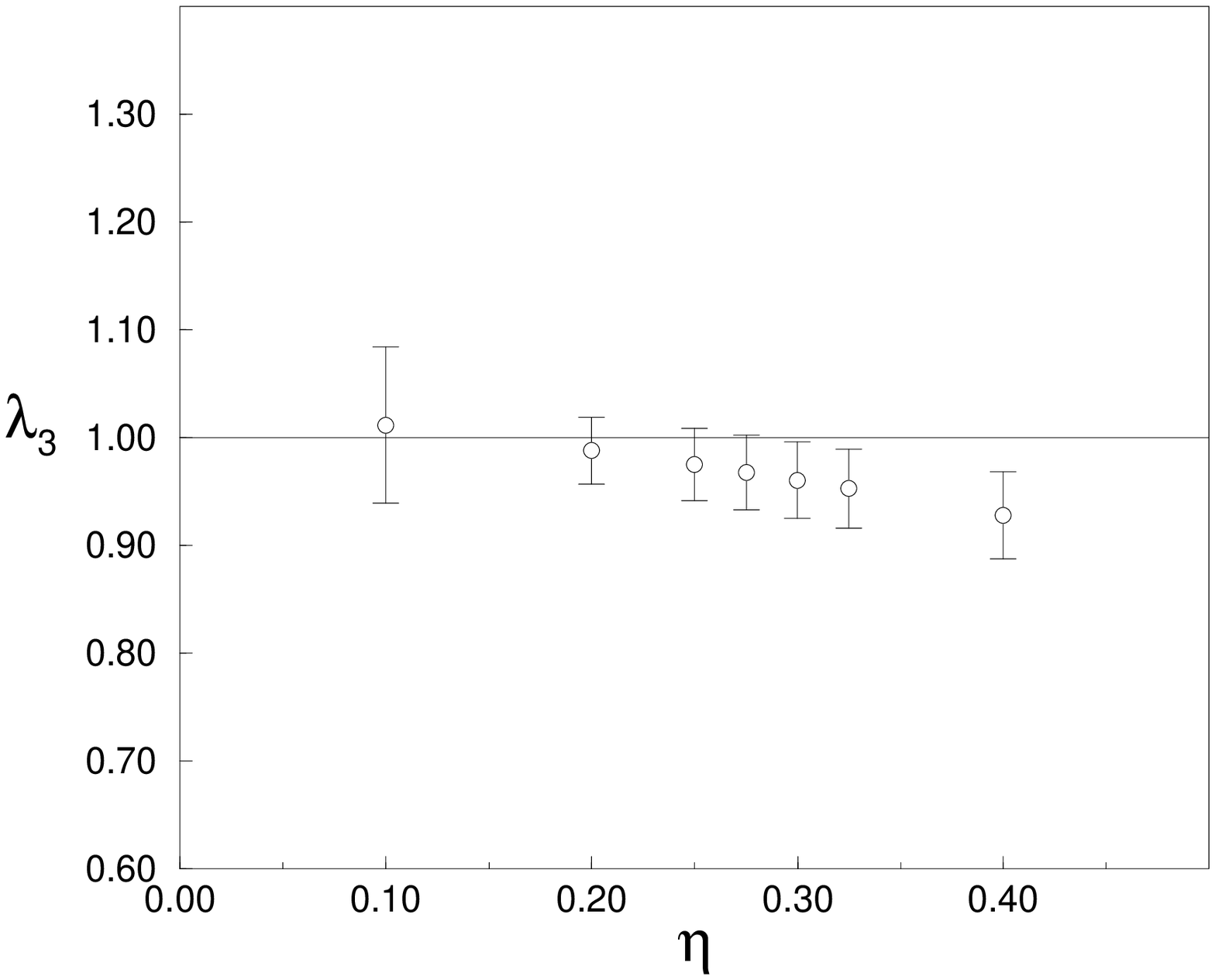,width=3.6 in}}
\caption{Third eigenvalue as a function of the anomalous dimension
at $a_W=16$, final lattice is $8^2$.}
\label{fig__third_eta}
\end{figure}

In fig.~\ref{fig__first_eta} we plot the first eigenvalue as a function
of the rescaling parameter. The first eigenvalue is very sensitive to
the value of the anomalous dimension. Statistical errors are relatively
small, certainly very small when compared with the error bars in the
the couplings. The first eigenvalue for the interval
of anomalous dimensions considered is
\bea\label{first__eigen}
\eta&=&0.25   \ , \  \lambda_1=2.015(7)
\nonumber\\
\eta&=&0.275   \ , \  \lambda_1=1.996(7)
\nonumber\\
\eta&=&0.30   \ , \  \lambda_1=1.976(7) \ ,
\eea
where the error is just statistical. Results are obtained after 3 RGTs, in
a $8^2$ lattice, so that finite size effect errors are smaller than the
statistical ones, as will be discussed.

In fig.~\ref{fig__third_eta} the third eigenvalue is plotted and as
already pointed out, it is compatible with one.

We can compute the FP action. For consistency with our analysis,
we will quote the results at $\eta=.30$, although given the relative large
error bars, it is essentially insensitive to $\eta$ within the  
range $(0.25,0.30)$.

\begin{table}[htb]
\centerline{
\begin{tabular}{|c|l||c|l||c|l|}
\multicolumn{1}{c}{Operator} & \multicolumn{1}{c}{$K$} & 
\multicolumn{1}{c}{Operator} & \multicolumn{1}{c}{$K$} & 
\multicolumn{1}{c}{Operator} & \multicolumn{1}{c}{$K$}      \\\hline
 1 & 0.16(12)   & 8  & 0.06(4)  & 24  & -0.017(5)  \\\hline 
 2 & 0.99(8)    & 9  & 0.05(2)  & 25  & -0.007(3) \\\hline
 3 & -0.22(3)   & 11 & 0.07(4)  & 27  &  0.005(2) \\\hline
 4 & 0.03(5)    & 12 & -0.06(2) & 33  &  0.029(5)\\\hline
 5 & -0.0016(3) & 15 & 0.12(2)  & 37  & -0.012(4) \\\hline
 6 & -0.90(5)   & 16 & 0.03(1)  &     &\\\hline
 7  & -0.17(4)  & 21 & 0.03(1)  &     &\\\hline
\end{tabular}}
\caption{FP Hamiltonian at $a_W=16$. Only the non-vanishing couplings are 
         displayed.}
\label{tab__FP}
\end{table}

\subsection{The different sources of errors}

We now analyze the different sources of errors that may appear. We treat
each case separatedly.

\subsubsection{error e(1)} 

This is the systematic error that may appear if some operators having
sizeable coupling are not considered into Eq.~\ref{def_matrices}. 
From previous results, we obtained that FP couplings decay at least 
exponentially, both with length and distance, and as we parametrized 
the effective interactions generated along the flow within this 
assumption, we should not expect large systematic errors coming 
from truncation effects.

In the case of the evaluation of couplings, a typical plot of the 
evolution in the value of the coupling as more operators are 
considered is shown in fig.~\ref{fig__trunc_coup6}. Results are 
strongly sensitive to the inclusion of higher dimension operators 
but rather mildly to length. Furthermore, statistical error bars
grow considerably with the inclusion of more operators. Eventually
there is flat plateau, but only after enough number of operators are
included. If we assume that the systematic error coming from the
truncation, consistently with previous assumptins, decays exponentially,
we estimate that is smaller than the statistical error. Although the
truncation error is not completely negligible, statistical errors
should be diminished to make it apparent. In other words, if larger
statistics were available, higher dimensional operators should
be considered.

Concerning the evolution of eigenvalues, we do find a flat plateau
which sets in with the inclusion of relatively few operators, see
fig.~\ref{fig__trunc_eigen1} and fig.~\ref{fig__trunc_eigen2}. 
For the case of eigenvalues we cannot find any evidence for a 
significant systematic error coming from the truncation.

We conclude that the operators incuded are 
consistent within the statistics we considered. Errors coming from the
truncation are smaller than the statistical one.

\begin{figure}[p]
\centerline{\epsfig{file=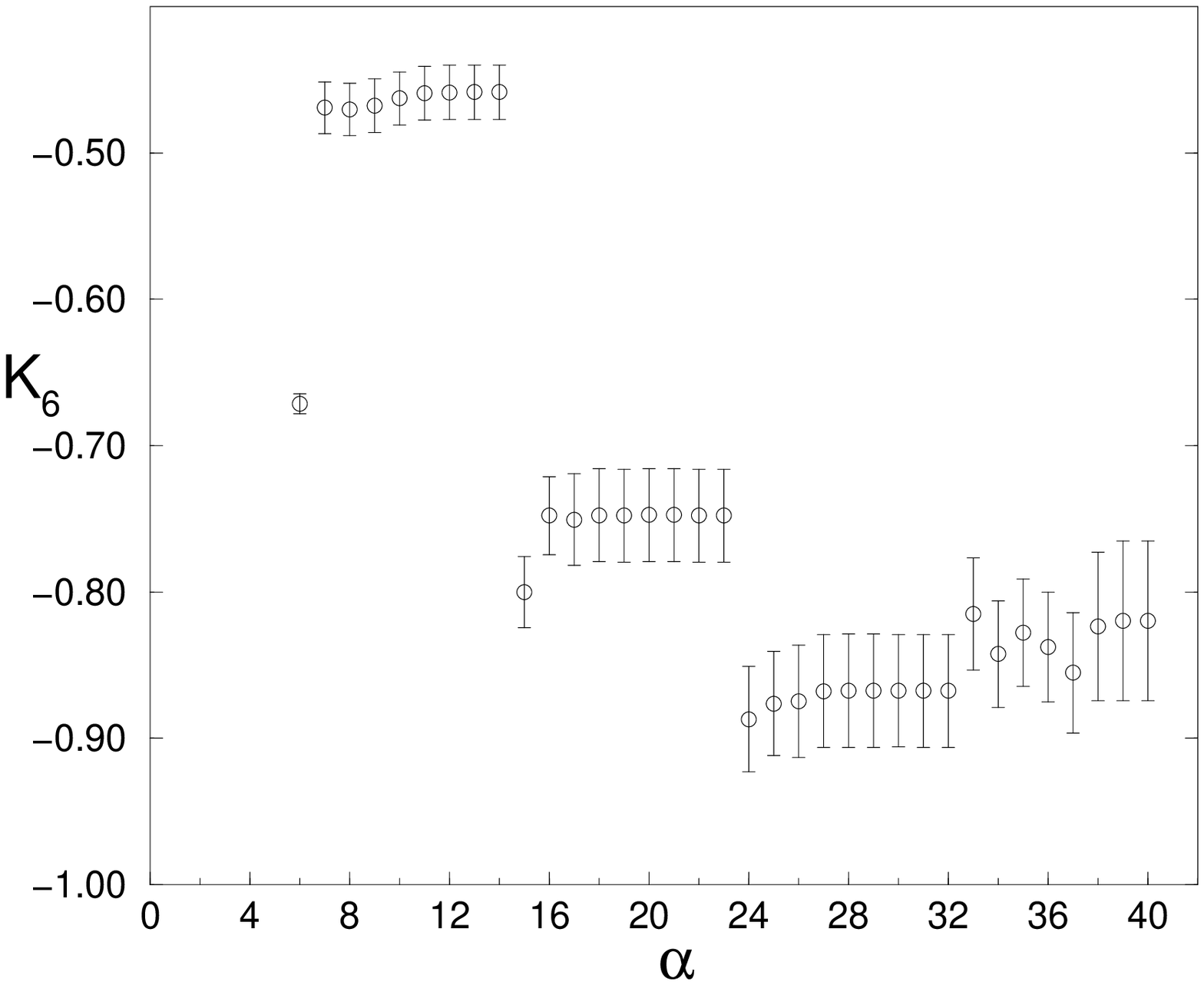,width=2.5 in}}
\caption{Evolution of coupling 6 as a function of operators included.}
\label{fig__trunc_coup6}

\centerline{\epsfig{file=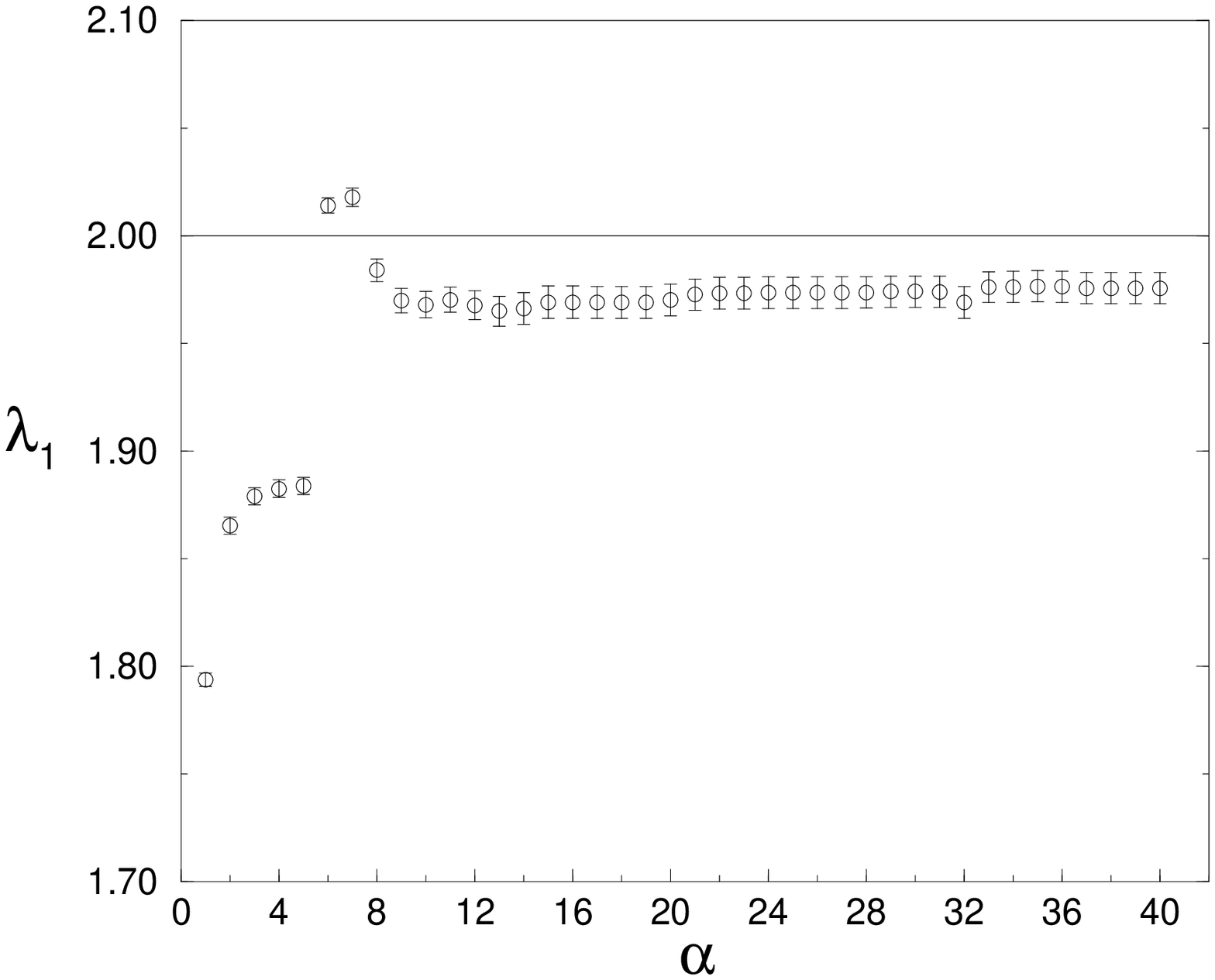,width=2.3 in}}
\caption{Evolution of the first eigevalue as a function of operators
included.}
\label{fig__trunc_eigen1}

\centerline{\epsfig{file=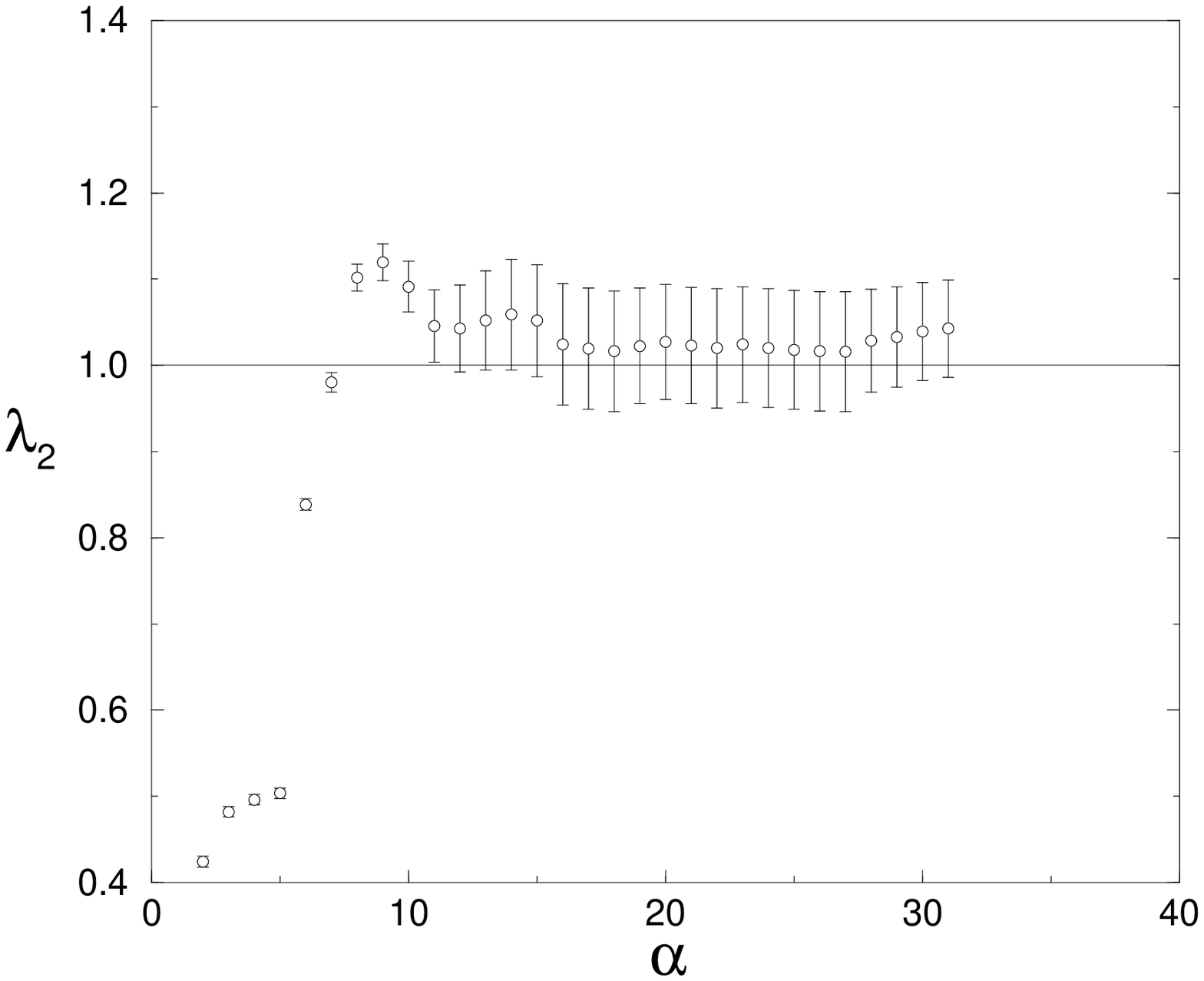,width=2.3 in }}
\caption{Evolution of the second eigenvalue as a function of operators
included.}
\label{fig__trunc_eigen2}

\end{figure}

\subsubsection{error e(2)}

Finite size effects enter into the game when there are operators with
sizeable coupling constants whose length is greater than the linear size of 
the lattice, $L$.  Both the couplings and eigenvalues will then acquire 
a dependence on $L$. Of course this error should be more 
apparent as the lattice is smaller. Our results at a $4^2$ do show
finite size effects, but as our main conclusions are drawn from 
the analysis of a $8^2$ lattice, we analyzed the effect in those.
In fig.~\ref{fig__error2} we compare the couplings of the effective 
interaction generated after 2 RGTs starting from a $32^2$ and $64^2$
lattices respectively. If there are finite size effects, these 
couplings should differ as they have different $L$ dependences, 
irrespective of how we truncated the expansion (we truncate it in the same
way in both lattices), and if we are reaching a FP( this is a question 
addressed by comparing successive RGTs). From fig.~\ref{fig__error2} we
cannot detect a significant dependence on $L$ within statistical errors. 
We conclude then, that finite size errors are within
statistical ones if the smallest lattice used is a $8^2$ one.

\begin{figure}[htb]
\centerline{\epsfig{file=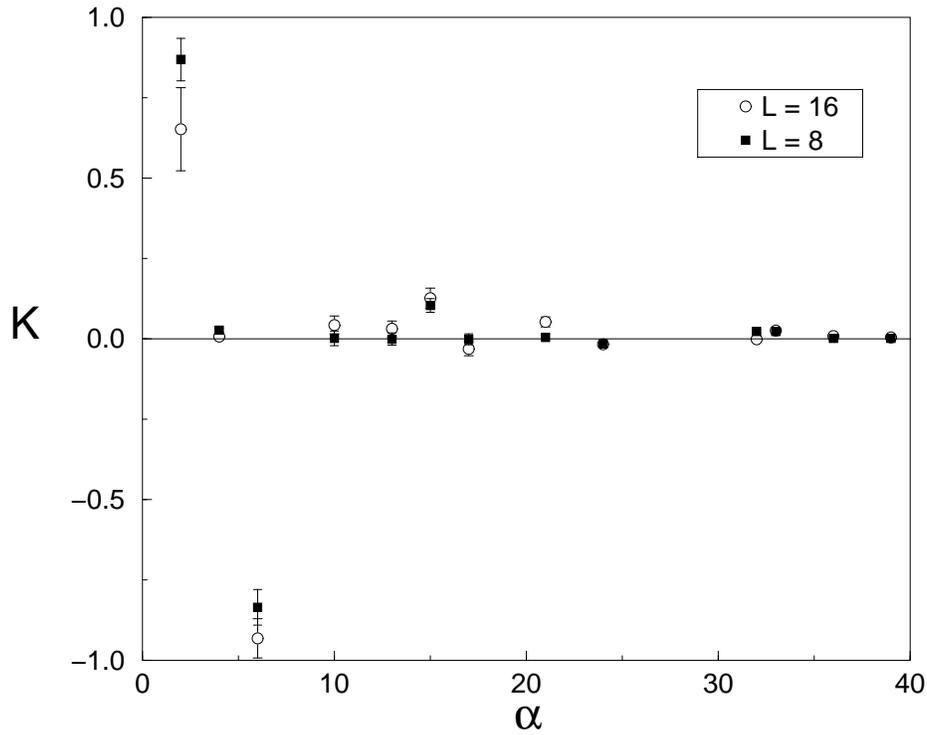,width=\textwidth}}
\caption{Comparison of couplings at $a_W=16$, $\eta=.275$ computed after
2 RGTs starting from a $64^2$ and $32^2$ lattice respectively.}
\label{fig__error2}
\end{figure}

\subsubsection{error e(3)} 

If the canonical surface is not exactly at criticality, the right infrared
FP cannot be reached. The RG trajectories eventually flow towards the 
renormalized trajectory instead, and the matching procedure
should eventually deteriorate. Unfortunately, we cannot perform
more RGTs to check if this is the case, and we do not have a more
precise estimate of the critical couplings to compare with the results
already obtained.
From the matching criterion (precisely the method more sensitive to this 
error) $\eta=.30$ is clearly preferred and the matching criterion 
is just marginally consistent with the exact result $\eta=0.25$.
This  makes us
suspect that there is a small systematic error of this kind in
our calculations, which unfortunately we cannot estimate. 

\subsubsection{error e(4)} 

This error is associated with the transformation, and it is not peculiar
to a finite lattice approximation. It appears when  
couplings in eigenvectors fail to converge. 

For the sake of clarity, 
in the appendix sect.~\ref{app__nonconv} we present a simple model
that can be solved exactly (the Gaussian model), where an explicit 
example of this problem is shown. 

An obvious criterion to detect this problem seems to compute the 
eigenvectors, diagonalizing the ${\rm T}$ matrix, explicitly  
checking that they decrease as a function of
length and dimension. For the first three eigenvalues 
(the $2$ and the two $1$s)
that we already computed we do not find a decay of the eigenvector components
as apparent as it was with the FP couplings. 
The next two eigenvalues at $a_W=16$ and 
$\eta \in (0.25,0.30)$ form a complex pair, so we think that those 
eigenvalues are affected by error e(4). As the eigenvectors are complex
as well, we do not know how to extract any consequence from the analysis
of the coefficients. In any case, let us mention that finite lattice 
approximations to simpler models do also show complex pairs when the
eigenvectors fail to converge, so it may be the case here as well. 
More statistics should clarify this issue.

\section{Conclusions}\label{sect__conc}

\subsection{Summary of the results}

In this paper we studied finite lattice approximations in models
of unconstrained spins. We discussed the properties that a 
renormalization group transformation should have in order that a finite
lattice approximation may be organized into a controlled and systematic 
expansion, able to give accurate and rigorous results. We also discussed the 
difficulties that appear in computing the anomalous 
dimension of the field.

As a non-trivial example, we studied the linear sigma model in two
dimensions in the domain of attraction of the Ising fixed point, using
the Bell-Wilson RGT \cite{BELL1}, which has a free parameter $a_W$.
The properties of the RGT depend on $a_W$ and the goal was to 
properly identify the values of $a_W$ that optimally fullfill the 
two important properties relevant for us, the short-rangeness of
the effective interactions generated and a rapid approach to the
Ising fixed point.

To address the issue of how short-ranged are the effective interactions
essential use has been made of the Schwinger-Dyson equations 
Eq.~\ref{def_coup} criterion 1(a). Overall the method performs well, but 
the statistical error bars are relatively large, which in some cases limits 
the accuracy of the results. In any case, we have provided convincing 
evidence that the decay ansatz Eq.~\ref{dec_FP_value} is well satisfied 
(see fig.~\ref{fig__type0dis}, fig.~\ref{fig__type1dis} and
fig.~\ref{fig__type2}), which in turn provides
the justification for the expansion we propose for the effective
interactions. Our results clearly show that the smaller the value 
of $a_W$ the better the decay properties and that decay with dimension
is slower than with distance. If more statistics would be available,
the expansion should include more higher dimensional operators than 
higher length.

Although the smaller $a_W$, the more short-ranged the fixed point action 
is, we found that for slightly higher values ($a_W\sim16$) the canonical
surface is closer to the corresponding FP. To reach this conclusion 
we computed the couplings generated after successive RGTs and checked
whether they agree, criterion 2(a). This was helpful to rule out 
high values of $a_W$, and to provide strong evidence for a FP at 
$a_W=16$($\eta=0.25$) (see fig.~\ref{fig__eta25__aw16match}).
We also used the matching criterion of observables
2(b) and found only a tiny dependence on $a_W$, enough to show that 
$a_W=16$ is the optimal case (see fig.~\ref{fig__aw_sensitivity}).

Having identified the value $a_W=16$ as the best suited for a finite 
lattice approximation, we computed the rescaling of the field, and
found an anomalous dimension $\eta \in (0.25,0.30)$. Although compatible 
with the exact result $\eta=0.25$, our results favor values closer
to $\eta=.30$. The first criterion used to address this issue, the
eigenvalue being one, led to a window of acceptable anomalous dimension 
$(0.25,0.30)$ after a further assumption that there are 2 marginal 
directions. This further assumption comes as a particularity of the 
two-dimensional linear sigma model and should not show up in other 
models. The matching criterion 3(b), which turned out
to be a very sensitive criterion (see fig.~\ref{fig__match_two}),
clearly favored $\eta \sim 0.30$.

After all this previous study, we present the final results for the
critical exponents,
\bea\label{final_exp}
\eta&\in&(0.25,0.30)
\\
\nu&=&1.00(2)
\\
\lambda_2&=&1.0(4)
\\
\lambda_3&=&1.0(4) \ ,
\eea
where the error merely indicates the uncertainty in the determination of the 
anomalous dimension, which in turn, depends on the statistical error and
the lack of criticality, as explained. We would like to emphasize that
all the rest of systematic errors have been considered and shown to be
smaller (see the discussion at the end of the last section).
We also computed the FP Hamiltonian and found that it may be 
parametrized with a few operators (see table~\ref{tab__FP}).
Finally, we provide arguments that within a Bell-Wilson RGT one can
only compute a finite number of exponents, the rest are not accessible 
to numerical methods due to error e(4). Most likely, this number reduces 
the accessible eigenvalues to the ones we quote, although it is possible 
that this is just a reflection of the modest statistics used. 

\subsection{Perspective and outlook}

By optimizing the transformation used
we have been able to get a great deal of information, including precise 
calculations for critical exponents, FPs, etc. with modest statistics,
and more importantly, to have under control the different source of
errors. There are, however, some problems that we should emphasize.
Concerning the determination of the anomalous dimension,
our results favor values of $\eta \sim .30$, although
$\eta=0.25$ is still acceptable. Higher statistics and a more accurate
determination of the critical surface should resolve this issue. Another
important point concerns real space renormalization group methods as they
seem to present a limitation concerning the number of, generally
irrelevant, critical exponents accessible within finite lattice 
approximations. Deeper understanding of the properties of RGTs are
necessary with the ultimate goal of proposing RGTs with better convergence 
properties and good short-rangeness behaviour, as is discussed in
the final appendix.

There are similar MCRG calculations for models of 
unconstrained spins using momentum RGTs \cite{JO}. As discussed,
RGTs in momentum space have different properties and are not directly 
comparable. Closer works to compare ours with are MCRG calculation  
for 3d Ising models using real space RGTs \cite{PAW1}, \cite{GUP1} 
and the most recent update \cite{GUP2}. Although the large statistics 
and big lattices reported in those papers are completely unattainable
with our resources, at a more conceptual level there are interesting
analogies we can draw. The RGT transformation used in those works is a 
majority rule, which is the spin-constrained version of the Bell-Wilson RGT
at $a_W=\infty$. We have seen that this is a very unfavorable case
concerning the short-ranged properties of the effective interactions
and the approach to the FP, so it may be the case as well for the 
constrained model (which can be obtained as the 
$\lambda \rightarrow \infty$ case of the unconstrained one). Let us recall
that this is not at odds with the good matching for expectation values
of operators. We find a surprisingly accurate matching for 3 RGTS even at 
$a_W=\infty$, despite that other criteria shows that we do not reach a 
FP after 3RGTs. Our best value for the matching singles out an anomalous
dimension slightly off from its exact value. Similarly, in 3d Ising MCRG 
calculations the anomalous dimension comes slightly off from
its more accepted value \cite{ZINN}. Secondly, only two even parity 
eigenvalues may be computed, a result which seems in agreement with the 
difficulties mentioned concerning the convergence of higher irrelevant 
eigenvalues. It would be interesting to perform similar calculations
using more elaborate RGTs.

Although, as discussed, some questions demand further clarification, we
have shown that with a suitable knowledge of the RGT, finite lattice 
approximations provide a systematic expansion to perform
calculations in the RG. We hope that this paper will help to arouse
and inspire further work in this field, which will allow
application of these methods to more general models.

\newpage

\appendix
\section{The Strong coupling expansion}\label{app__strong}

    We determine the CS by means of a
    high temperature expansion (HTE) of the susceptibility,
    \be\label{susc_exp}
    \chi=\sum_{i=0}^{\infty} \chi_1^{(i)}\kappa^i,
    \ee   
    where $ \chi_1^{(i)}$ depend on $\lambda$ through integrals like
    $I_{n}(\lambda)=\int_0^{\infty}d\phi\phi^n e^{-S[\phi]}$.
    Baker and Kincaid \cite{BAK1}, have tabulated its first 11 coefficients,
    and in such a case, only even integrals up to $n=12$ 
    have to be considered.

    Had we computed the whole series, the critical coupling is obtained 
    from the ratio method \cite{LEB1},
    \be\label{ratio}
    \kappa_c(\lambda)=\lim_{n \rightarrow \infty} r_n , \\\
	 r_n \equiv \chi_1^{(n-1)}/ \chi_1^{(n)},
    \ee
    but in our case must be extrapolated from finite n ratios. In such
    a case corrections to the preceeding formulae are correlated with
    the form of the closest singularity of the susceptibility to the 
    origin in the complex $\kappa$ plane. It is assumed
    that, 
    \be\label{ratio_scal}
    r_n=\kappa_c(\lambda)(1+\frac{1-\gamma}{n}+\sum_{j}\frac{c_{j}}{n^j}),
    \ee
    plus some exponential corrections $\exp(-na)$ arising from singularities
    farther, provided that within the disk 
    $|\kappa|=\kappa_c(\lambda)$ there are no further singularities. This
    is, unfortunately not the case for a $PSQ$ (plane square) lattice 
    (as it is always for loose-packed lattices),
    due to the presence of an antiferromagnetic singularity at
    $\kappa=-\kappa_c(\lambda)$, which brings in 
    nonanalytical oscillating power law corrections in Eq.\ \ref{ratio_scal},
    of the type,
    \be\label{ratio_scal_corr}
    r_n \sim \frac{{(-1)}^n}{n^{\gamma+\Phi}},
    \ee
    $\Phi$ is the critical exponent associated to the antiferromagnetic 
    singularity.

    The ratios in Eq.\ \ref{ratio}, of the expansion in Eq.\ \ref{susc_exp},
    are nearly constant with a monotonic decreasing oscillating trend.
    Using a double ratio method \cite{LEB1}, a rough estimate of the
    critical coupling may be obtained $\kappa_{re}$. We define then a new
    variable through,
    \be\label{new_var}
    z=\frac{2\kappa}{1+\frac{\kappa}{\kappa_{re}}},
    \ee
    so that we send the antiferromagnetic singularity almost 
    to $\infty$, and thus the susceptibility expressed in this new variables
    eliminates the oscillating term in Eq.\ \ref{ratio_scal_corr}
    as it picks up a large exponential correction.

    We can extrapolate now by using Eq.\ \ref{ratio_scal} for $n \geq 5$
    and we estimate the error from comparing extrapolations including 
    different invers powers in $n$. We also analyzed the ratio method if
    insted of $\kappa_{re}$, we input the new better estimate, but
    nothing is gained within error bars. In any case,
    Just to compare with known results, we obtained for $\lambda=\infty$
    (Ising limit),
    \be
    K_c=0.4407(2),
    \ee
    in agreement with the exact result $K_c=0.4406868$.

\section{The Non-convergence of eigenoperators}\label{app__nonconv}

In this appendix we show in an explicit example how exact eigenoperators
of the ${\rm T}$-matrix may not converge. We explictly show as an
example the Gaussian Model, defined as the most general 
translational-invariant quadratic Hamiltonian
\be\label{Gauss_model}
{\cal H}=\frac{1}{2}\sum_{n}\sum_{r} \rho(r)\phi(n)\phi(n+r) \ ,
\ee
where the sum is over sites of the corresponding lattice.
Under $k$-iterations of a linear RGT the model transforms into
itself,
\be\label{Gauss_model_ren}
{\cal H}_k=\frac{1}{2}\sum_{n}\sum_{r} \rho_k(r)\vartheta(n)\vartheta(n+r)
\ .
\ee
It is possible to derive an exact RG-equation for the 
RGT used in this paper \cite{BELL1}, which is not difficult
to generalize to a completely general linear RGT \cite{THESIS}.
We can actually solve the equations and find the exact FP and 
eigenoperators. We just quote the final result for 
the $j$-eigenoperator \cite{THESIS}
\be\label{j_eigenoperators}
\delta_{j}\rho(p)=\rho^*(p)^2 g_j(p) \ , \
g_{j}(p)=\sum_{l=-\infty}^{+\infty}\frac{1}{(p+2\pi l)^{(4-2j)}}
\prod_{\mu=1}^d g(p_{\mu}+2 \pi l_{\mu}) \ ,
\ee
where the equation is expressed in momentum space, $j=0,1,\cdots$,  
the dimension of the system is $d$ ($d=2$ in our case), 
and $\rho^*$ is the FP value of the 
couplings which is not directly relevant for our problem and will 
appear elsewhere. The function
$g(x)$ depends on the RGT. For the Bell-Wilson RGT Eq.~\ref{Bell_transf}
we have \cite{BELL1}
\be\label{g_Bell}
g(x)=\frac{\sin^2(\frac{x}{2})}{(\frac{x}{2})^2} \ .
\ee
In this case, eigenoperators Eq.~\ref{j_eigenoperators} are well defined
for $j=0$ (relevant), $j=1$ (marginal), $j=2$ (irrelevant), but the sum
in Eq.~\ref{j_eigenoperators} fails to converge for $j > 2$. Any numerical
method will necessary fail to get this eigenoperator right, as it has
just a formal meaning. This conforms very well to the numerical results 
reported in \cite{BELL1}.

To remedy this situation, one has to consider other RGTs, which 
have overlapping cells (see fig.~\ref{fig__overlap}). Let us mention
that these transformations have appeared in the context of perfect
actions \cite{HAS1}. For such a case one gets a g-function
\be\label{g_overlap}
g(x)=\frac{\sin^4(x)}{(\frac{x}{2})^4} \ ,
\ee
so now, for example, from Eq.~\ref{j_eigenoperators} the $j=3$ 
eigenoperator does converge, opposite to what it happens with 
Bell-Wilson RGT. In other words, the eigenvalue with $j=3$ becomes now 
accessible to numerical methods.
In fact, the RGT in fig.~\ref{fig__overlap} is just the simplest case
of an overlapping RGT. One may define more elaborated RGTs so 
that the $g-$function Eq.~\ref{g_Bell} becomes more convergent 
\cite{THESIS}.

\begin{figure}[htb]
\centerline{\epsfig{file=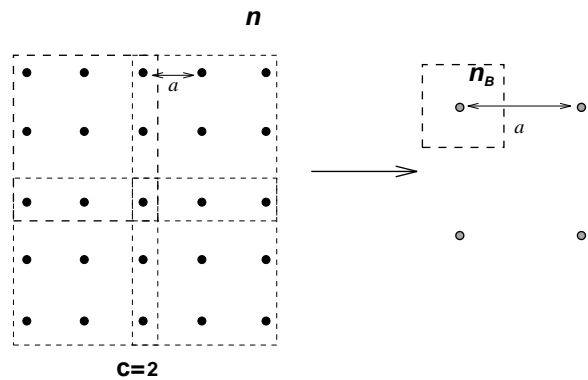, width=3 in}}
\caption{The simplest case of overlapping RGT.}
\label{fig__overlap}
\end{figure}

Let us recall that RGTs in momentum space are an extreme case of
overlapping RGTs. In such a case, it is easy to prove that eigenoperators
are perfectly convergent for any value of $j$.
However, this is a very unfavorable case concerning
the short-rangeness of the effective actions generated, because the 
decay ansatz Eq.~\ref{dec_FP_value} gets replaced by an algebraic
decay. It seems then, that there is some kind of uncertainty principle
concerning RGTs; One may have very well behaved RGTs concerning how 
short-ranged
the effective interactions are at the expense of some eigenoperators
having just a formal meaning, or one may choose RGTs having very well 
defined set of eigenoperators at the expense of having poor short-ranged 
behaviour of the Hamiltonian generated along the RG flow.

\bigskip
\bigskip
\bigskip

{\bf ACKNOWLEDGEMENTS}

We acknowledge M. Bowick for a careful reading of the manuscript.
We also acknowledge interest and discussions with J. Comellas, D. Espriu, 
P. Hasenfratz and R. Toral. The research of A.C., E.G. and A.T. was
supported by the U.S. Department of Energy under Contract No.
DE-FG02-85ER40237.

\end{document}